\documentclass[prl,aps,showpacs,twocolumn,preprintnumbers,amsmath,amssymb,superscriptaddress,nofootinbib]{revtex4-1}
\usepackage{amsmath,amssymb,amsfonts}
\usepackage{graphicx}
\usepackage[colorlinks=True,linkcolor=black,citecolor=blue,urlcolor=blue]{hyperref}
\usepackage{xcolor}
\usepackage{chngcntr}
\usepackage{braket}
\usepackage{hyperref}
\usepackage{nicefrac}
\usepackage{bm}
\usepackage{float}
\usepackage{ulem}

\usepackage{booktabs}

\newcommand{\red}[1]{{\textcolor{black}{#1}}}

\begin{document}

\title{Intrinsic and extrinsic photogalvanic effects in twisted bilayer graphene}

\author{Fernando Peñaranda}
\affiliation{Donostia International Physics Center, P. Manuel de Lardizabal 4, 20018 Donostia-San Sebastian, Spain}
\author{Héctor Ochoa}
\affiliation{Department of Physics, Columbia University, New York, NY 10027, USA}
\author{Fernando de Juan}
\affiliation{Donostia International Physics Center, P. Manuel de Lardizabal 4, 20018 Donostia-San Sebastian, Spain}
\affiliation{IKERBASQUE, Basque Foundation for Science, Maria Diaz de Haro 3, 48013 Bilbao, Spain}

\date{\today}
\begin{abstract}
The chiral lattice structure of twisted bilayer graphene with $D_6$ symmetry allows for intrinsic photogalvanic effects only at off-normal incidence, while additional extrinsic effects are known to be induced by a substrate or a gate potential. In this work, we first compute the intrinsic effects and show they reverse sign at the magic angle, revealing a band inversion at the $\Gamma$ point. We next consider different extrinsic effects, showing how they can be used to track the strengths of the substrate coupling or \red{electric} displacement field. We also show that the approximate particle-hole symmetry implies stringent constraints on the chemical potential dependence of all photocurrents. A detailed comparison of intrinsic vs. extrinsic photocurrents therefore reveals a wealth of information about the band structure and can also serve as a benchmark to constrain the symmetry breaking patterns of correlated states. 
\end{abstract}
\maketitle

\emph{Introduction} - Magic-angle twisted bilayer graphene~\cite{Bistritzer11} (TBG) is a unique material platform where the combination of {  non-localized} flat bands at the Fermi level and strong correlations give rise to a variety of insulating ground states~\cite{Cao2018Insulators} and unconventional superconductivity~\cite{Cao18SC}. {   The flat bands hosting these states do not originate from trivial isolated orbitals but display instead complex wavefunctions, so the challenge of understanding the insulating and superconducting states would greatly benefit form experiments that directly probe the quantum geometry and symmetries of such wavefunctions}.  

Interestingly, a set of such experimental probes is enabled by the chiral lattice structure of moir\'e lattices~\cite{Kim16} like TBG, where all mirrors and inversion are broken. These include non-linear optics like photogalvanic effects~\cite{Liu20Anomalous,Kaplan22,Chaudhary22} and second harmonic generation~\cite{Zhang22Correlated,Yang20}, and spatially dispersive linear optics like optical activity and circular dichroism \cite{Morell17,Stauber18,Stauber20,Wang20Optical,Chang22}. All of these probes are known to be uniquely sensitive to different aspects of wavefunction geometry, and have been proposed to uncover many interesting properties of the TBG bands. The photogalvanic effects in particular, where a DC current is generated by light, are given by
\begin{align}
J_i = \sigma_{ijk} (E_jE_k^* + E_kE_j^*) + \eta_{ijk} (E_jE_k^* - E_kE_j^*),     \label{mainPGE}
\end{align}
where $\sigma_{ijk}$ is real and symmetric, $\sigma_{ijk}= \sigma_{ikj}$, and encodes the linear photogalvanic effect (LPGE), while $\eta_{ijk}$ is imaginary and antisymmetric, $\eta_{ijk}=-\eta_{ikj}$, and encodes the circular photogalvanic effect (CPGE).
\begin{figure}[t]
    \centering
\includegraphics[width=0.45\textwidth]{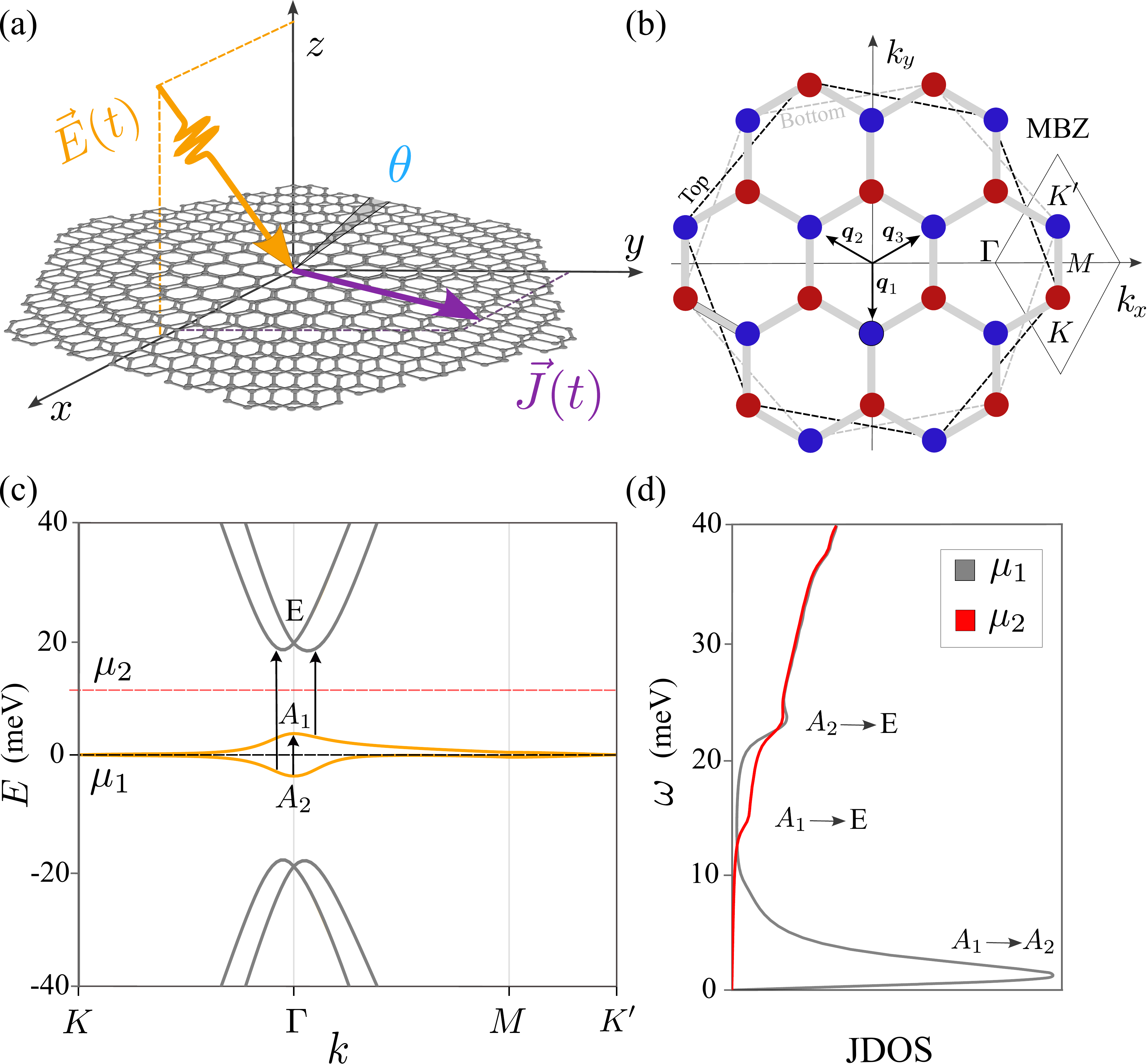}
    \caption{(a) Sketch of a TBG flake under out-of-plane EM radiation. An off-normal incident field $\boldsymbol E = E_i \boldsymbol e_i$ with $E_z \neq 0$ and associated second-order DC current $\boldsymbol J$ are shown in yellow and purple, respectively. (b) Brillouin zone of TBG. The moir\'e Brillouin zone is highlighted in black. Black and gray dashed hexagons refer to the first Brillouin zone of top and bottom graphene monolayers, respectively. (c) Band structure of the $\nu=+1$ valley of TBG Hamiltonian in Eq. \eqref{continuummodel} with $\theta = 1.05^\circ$. Bands are labeled according to their $D_3$ irreducible representation at $\Gamma$. (d) Joint density of states \red{for vertical interband transitions} at the two $\mu$ values indicated by dashed lines in (c), with shaded regions representing involved transitions as a function of $\omega$.}
    \label{Fig1}
\end{figure}
For TBG in its high-temperature state with \red{$D_6$ symmetry generated by $C_{6z}$ and $C_{2x}$ (see Fig.~\ref{Fig1}~a~and~b for conventions}),
the only allowed PGE components are $\sigma_{xyz} = - \sigma_{yzx}$ and $\eta_{xyz} = \eta_{yzx}$, involving an out of plane electric field $E_z$ which requires off-normal incidence to be measured (see Fig.~\ref{Fig1}). Off-normal photocurrents are commonly measured in experiment~\cite{Yuan14,Quereda18,Ni21} but not often calculated for 2D materials because the coupling of the electric field has to be implemented differently for the in-plane periodic directions and the ouf of plane finite one ~\cite{Wang20Optical,Chen23Crossed,Zheng2023}. Previous predictions in TBG have thus focused on the normal incidence components induced by symmetry breaking. For example, BN encapsulation reduces the symmetry to $C_{3z}$ and enables $\sigma_{xxx}$ and $\sigma_{yyy}$,  \red{among others} \cite{Liu20Anomalous,Kaplan22,Zhang22Correlated,Chaudhary22}, and breaking $C_{3z}$ with strain \red{enables $\eta_{yxy}$} \cite{Arora21}.
Experimentally, optical spectroscopy near the magic angle is already feasible~\cite{Hesp21,Sunku21} and photogalvanic currents have been observed at normal incidence~\cite{Otteneder20,Hubmann22,Ma22}.

Knowledge of the off-normal incidence photocurrents is however important as they probe the intrinsic properties of the TBG Bloch bands, rather than the effects of symmetry breaking, be it external or correlation-induced. In this work, we present a detailed calculation of the intrinsic, off-normal photocurrents, and show that they change sign at the magic angle, both at neutrality and finite chemical potential. We also compute the extrinsic photocurrents due to symmetry breaking for comparison, and show that both intrinsic and extrinsic photocurrents satisfy stringent constraints when the chemical potential is reversed due to an approximate particle-hole symmetry (PHS). Finally, we evaluate the effect of weak PHS breaking and overall establish how these effects can be used to extract detailed information about the TBG band structure. 

\emph{Photocurrents in the continuum model} - We compute the non-linear optical responses with the length gauge Hamiltonian $H= H_0 + H_E$, where light couples to the position operator as $H_E = -e \vec r \cdot \vec E$, and $H_0$ describes the unperturbed electronic structure. We employ the continuum model~\cite{dosSantos12}, which describes the low-energy electronic states around the two valleys ($\nu=\pm 1$) of the two graphene layers when the bottom layer is rotated counterclockwise by an angle $\theta$ with respect to the top one. For small twist angles we assume that charge is conserved independently on each valley sector, $H_0=\sum_{\nu}H_{\nu}$,  and the matrix elements of $H_\nu$ in the layer basis are
\begin{align}
H_{\nu} = \begin{pmatrix}
    -i\hbar v_F\, \boldsymbol{\sigma}_{\nu} \cdot \boldsymbol{\partial}  & T_\nu(\boldsymbol r)\\
    T_\nu^\dagger(\boldsymbol r) & -i\hbar v_F\, \boldsymbol{\sigma}_{\nu} \cdot \boldsymbol{\partial} \\
\end{pmatrix}.
\label{continuummodel}
\end{align}
Here $\boldsymbol{\sigma}_{\pm}=(\pm \sigma_x,\sigma_y)$ is a vector of Pauli matrices acting on $(A,B)$-sublattice space and $v_F$ is the Fermi velocity. The coupling between layers is given by the moir\'e potential, $T_\nu(\boldsymbol r) =\sum_{n=1}^3T_{n, \nu} e^{-i \nu \boldsymbol q_n \cdot \boldsymbol r}$, where
\begin{align}
    T_{n+1, \nu} &= w_{AA} \sigma_0 + w_{AB} [\sigma_x \cos (\phi n) + \sigma_y \sin (\phi n\nu)],
\end{align}
$w_{AA}$ and $w_{AB}$ are the interlayer couplings between AA and AB stacking regions, respectively, $\phi = 2\pi/3$, and $\boldsymbol q_{1,2,3} = k_\theta \{ \left(0, -1 \right), \left(-\sqrt{3}/2, 1/2 \right), \left( \sqrt{3}/2, 1/2 \right) \}$ are the momentum boosts resulting from the relative shift $k_\theta = 8 \pi/(3a) \sin{\theta/2}$ of the three equivalent Dirac points of each graphene layer, see Fig.~\ref{Fig1}~b.

The band structure for valley $\nu=+1$ is shown in Fig. \ref{Fig1}~c with parameters $\hbar v_F/a = 2135.4$ meV, $\omega_{AA} = 79.0$ meV, and $\omega_{AB} = 97.5 $ meV at $\theta = 1.05^\circ$, which is below the magic angle  $\theta_M \approx 1.08^\circ$ for this set of parameters. A phenomenological broadening was set to $0.5$ meV in all simulations. 
Figure~\ref{Fig1}~d displays the joint density of states \red{for vertical interband transitions contributing to} 
resonant optical responses \red{at finite frequency}. 
At neutrality (black curve) transitions between the flat bands give rise to a first peak in $\omega$, followed by an onset due to transitions from lower flat band to dispersive bands. At full filling (red curve) the first process is blocked and a new onset appears due to transitions from upper flat band to dispersive bands. This pattern is followed by all responses computed below. 

\begin{figure}[t]
    \centering
\includegraphics[width=0.5\textwidth]{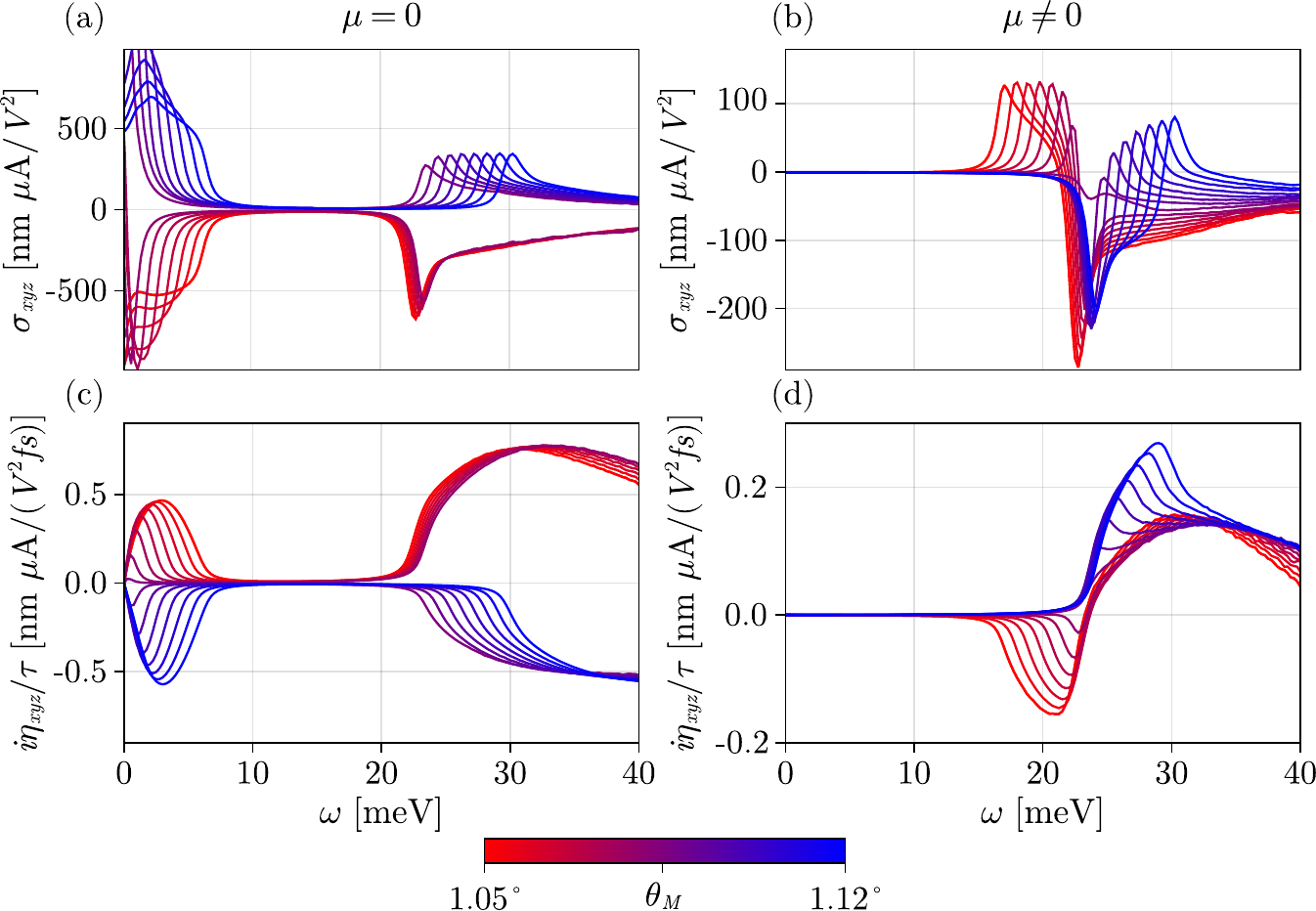}
    \caption{Intrinsic photogalvanic tensors at off-normal incidence. (a) Shift current $\sigma_{xyz}$ at neutrality ($\mu = \mu_1$ in Fig. \ref{Fig1}c). (b) $\sigma_{xyz}$ for fully filled flat bands ($\mu = \mu_2$). (c,d) Same for the injection current $\eta_{xyz}$. Only the case $\mu>0$ is shown in b) and d); $\mu<0$ is the same by PHS. Different colors refer to different $\theta$. A characteristic sign reversal when $\theta = \theta_M$ is observed in all cases.}
    \label{Fig2}
\end{figure}

\begin{figure*}[t]
    \centering
\includegraphics[width=1\textwidth]{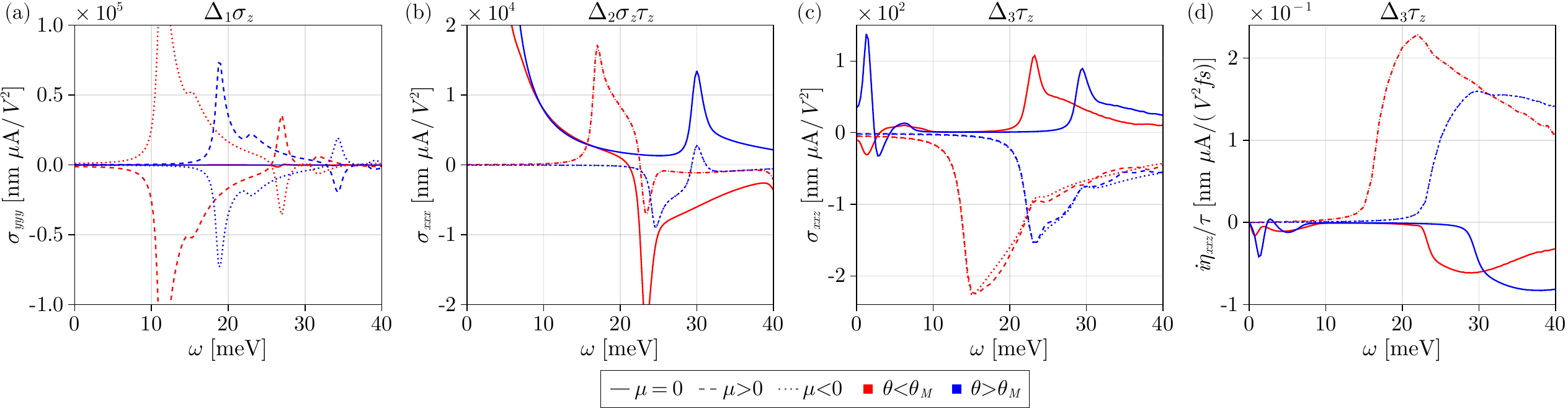}
    \caption{Photogalvanic components induced by the three symmetry-breaking perturbations in Table \ref{components}: (a) $\sigma_{yyy}$ with a $\Delta_1 \sigma_z$ term, (b) $\sigma_{xxx}$ with a $\Delta_2 \sigma_z \tau_z$ term, and (c) $\sigma_{xxz}$ and (d) $\eta_{xxz}$ with a $\Delta_3 \tau_z$ term. Red (blue) encodes $\theta<\theta_M$ ($\theta>\theta_M$), while solid, dashed, and dotted lines refer to $\mu_1$ and $\pm \mu_2$ in Fig. \ref{Fig1}c, respectively. $\Delta_1 = \Delta_2 = 10$ meV and $\Delta_3 =15 $ meV.}
    \label{Fig3}
\end{figure*}

We consider the photocurrent due to shift and injection current mechanisms, which in the presence of time-reversal symmetry $\mathcal{T}$ contribute only to the LPGE and CPGE, respectively~\cite{Aversa95,SM}. We diagonalize Eq. \ref{continuummodel} with Bloch eigenstates given by $H_0 \left|n \right> = E_n \left|n\right>$ ($\boldsymbol{k}$ dependence is implicit). The dipole coupling to light will require matrix elements of both in-plane and out-of-plane components of the position operator. The in-plane position matrix elements and their derivatives are evaluated in the standard way~\cite{Aversa95,Shkrebtii00}, while the $z$ component in a bilayer system simply becomes the operator $\hat{r}^z = \tau_z c/2 $ (where $\tau_i$ Pauli matrices act in layer space) with $c = 3.3\ \text{\AA}$  the interlayer distance \cite{Wang20Optical,Chen23Crossed,Zheng2023}. With these ingredients, the shift current has the expression
\begin{align}
\sigma_{ijk} = \frac{\pi e^3}{2\hbar^2} \int_{\boldsymbol{k}} \sum_{nm} f_{nm}[r^k_{nm;i}r^j_{mn}-r^k_{nm}r^j_{mn;i}] \delta(\omega - \omega_{nm}),
\end{align}
where $\int_{\boldsymbol{k}} =\int \tfrac{d^2k}{(2\pi)^2}$, $\omega_{nm} = (E_n-E_m)/\hbar$, $r_{nm}^i = \left<n\right| i\partial_{k_i} \left|m\right>$ for $i=x,y$ and $r^z_{nm} = \left<n\right| \hat{r}^z \left|m\right>$. \red{ $r^j_{nm;i}$ denotes the $i$'th in-plane component of the generalized derivative of $r^j_{nm}$,} computed with the aid of the standard sum rule for the in-plane position operators~\cite{Aversa95,Shkrebtii00} \red{(see also Eq.~19 in the Supplementary Material)}. For the $z$ component we obtain (see Ref. \cite{SM})
\begin{align}
r^z_{nm;i} &= \sum_{p\neq m} r^z_{np}\frac{v^i_{pm}}{\omega_{pm}} -  \sum_{p\neq n} \frac{v^i_{np}}{\omega_{np}}r^z_{pm},
\end{align}
where $v^i_{nm} = \left<n\right| \partial_{k_i}H_0 \left|m\right>$. The steady state injection current in the relaxation time approximation is proportional to the scattering time $\tau$ and given by
\begin{align}
\eta_{ijk} = \tau \frac{\pi e^3}{\hbar^2}\int_{\boldsymbol{k}} \sum_{nm} f_{nm} \partial_{k_i}\omega_{nm} r_{nm}^jr_{mn}^k \delta(\omega - \omega_{nm}).    
\end{align}
The figures will display the intrinsic quantity $\eta_{ijk}/\tau$. 
  
\emph{Symmetry constraints} - Next we consider how symmetries constrain the different photogalvanic tensors. We consider only in-plane currents and thus disregard $\sigma_{zij}$ and $\eta_{zij}$, which cannot be measured in a standard photocurrent experiment. As explained in the introduction, the point group $D_6$ of the continuum model in Eq. \ref{continuummodel} then only allows $\sigma_{xyz} = - \sigma_{yzx}$ and $\eta_{xyz} = \eta_{yzx}$, which we define as the intrinsic photocurrents. 

In the presence of a substrate or BN encapsulation, symmetries can be broken by three different terms in the Hamiltonian, $H_{\rm subs.} = \sum_i \Delta_i M_i$ with $i=1,2,3$, where $M_1 = \sigma_z$ is a sublattice potential that is the same in both layers, $M_2 = \sigma_z \tau_z$ is a sublattice potential that is opposite in each layer, $M_3 =\tau_z$ is an \red{electric} displacement field, and $\Delta_i$ represents the strength of each perturbation. Each perturbation \red{breaks} a different symmetry and 
allows different extra components of the photocurrent tensors, which we define as the extrinsic photocurrents. In particular, $\Delta_1$ allows $\sigma_{yyy}=-\sigma_{yxx}=\red{-}\sigma_{xxy}$, $\Delta_2$ allows $\sigma_{xxx}=-\sigma_{xyy}=-\sigma_{yxy}$ (both normal incidence photocurrents \cite{Liu20Anomalous,Kaplan22,Zhang22Correlated,Chaudhary22}) while $\Delta_3$ allows $\sigma_{xxz}=\sigma_{yyz}$ and $\eta_{xxz}=\eta_{yyz}$ (which require off-normal incidence). The symmetries preseved by each term and the components that it allows are summarized in Table  \ref{components} (symmetry related components are not shown).

\begin{table}
\centering
\begin{tabular}{|c||c|c|c|}
\hline
$H_{\rm subs.}$ & $ \Delta_1 \;\sigma_z$ & $\Delta_2 \; \sigma_z \tau_z$ & $\Delta_3 \; \tau_z$ \\ \hline
\red{Point symmetries} & $\red{C_{3z}}$, $C_{2y}$ & \red{$C_{3z}$}, $C_{2x}$ & $\red{C_{3z}}$, $C_{2z}$  \\ \hline
PH symmetries & $\mathcal{C}$ & $C_{2z}\mathcal{TC}$ & $\mathcal{C}$ \\ \hline
LPGE & $\sigma_{yyy}$ & $\sigma_{xxx}$ & $\sigma_{xxz}$\\
\hline
CPGE & none & none & $\eta_{xxz}$   \\ \hline
\end{tabular}
\caption{Extrinsic photocurrents beyond $\sigma_{xyz}$ and $\eta_{xyz}$ induced by symmetry breaking perturbations in $H_{\rm subs.}$, as explained in the text. From left to right, we consider layer-even sublattice potential $\Delta_1$, layer-odd sublattice potential $\Delta_2$ and an interlayer bias $\Delta_3$. The point group \red{generators} and particle-hole symmetries preserved by each perturbation are listed in the second and third row, respectively. 
}\label{components}
\end{table}

In addition to the previously discussed exact symmetries, the continuum model in Eq. \eqref{continuummodel} also has a PHS ~\cite{Moon13} $\mathcal{C}:$ $U_C H U_C = -H^*$ with $U_C = \sigma_x \tau_y$, known to lead to important consequences like selection rules for optics~\cite{Moon13,Morell17,Ahn21}, a stronger version of the Wannier obstruction~\cite{Song19,Song21}, and an enlarged symmetry of the projected Coulomb interaction~\cite{Bultinck20,Bernevig21}. As we now discuss, PHS also leads to crucial constraints for the photocurrent responses. These constraints are component dependent, as can be seen from the fact that the in-plane position operator $i \partial_{k_i}$ is invariant under $\mathcal{C}$, while the out of plane one $c\tau_z$ is odd. This leads to another main result of this work, proven in Ref. \cite{SM}, which is that normal incidence photocurrents change sign under $\mathcal{C}$, but oblique incidence photocurrents with an odd number of $z$ indices are invariant under $\mathcal{C}$. Since the chemical potential $\mu$ is odd under PHS, for intrinsic photocurrents we thus have 
\begin{align}
\sigma_{xyz}(\mu) = \sigma_{xyz}(-\mu) ,  & & \eta_{xyz}(\mu) = \eta_{xyz}(-\mu). \label{PHintrinsic}
\end{align}
Similar constraints are also applicable to extrinsic photocurrents as long as the perturbation that induces them also preserves $\mathcal{C}$ or its combination with another symmetry element (see third row of Table \ref{components}). In the presence of $\Delta_1$, which is invariant under $\mathcal{C}$, we have 
\begin{align}
\sigma_{yyy}(\mu) = -\sigma_{yyy}(-\mu),\label{PHSyyy}
\end{align}
while in the presence of $\Delta_2$, which is invariant under $\mathcal{C}C_{2z}\mathcal{T}$, we have 
\begin{align}
\sigma_{xxx}(\mu) = \sigma_{xxx}(-\mu),
\end{align}
because the shift current LPGE is odd under $C_{2z}\mathcal{T}$. Finally, in the presence of $\Delta_3$, again invariant under $\mathcal{C}$, we have 
\begin{align}
\sigma_{xxz}(\mu) = \sigma_{xxz}(-\mu) ,  & & \eta_{xxz}(\mu) = \eta_{xxz}(-\mu). \label{PHoffnormal}
\end{align}

\emph{Intrinsic photocurrents} - The computed intrinsic photocurrents of TBG as a function of the incident frequency are shown in Fig. \ref{Fig2}, for chemical potentials at neutrality and for full occupation of the flat bands and for a range of twist angles. We first note their sizable magnitudes: Shift currents range around 100-500 ${\rm nm \; \mu A /V^2}$, which is comparable to the largest values found in 2D materials~\cite{Sauer23}, and injection currents display similar magnitudes if we take $\tau$ as typical lifetimes from transport experiments, 100-200 fs \cite{Monteverde10}. Regarding the frequency dependence, in addition to the general JDOS features anticipated, we observe that both $\sigma_{xyz}$ and $\eta_{xyz}$ change sign as the magic angle is crossed, a main result of this work. This sign reversal can be tracked back to a band inversion at the $\Gamma$ point when crossing the magic angle, where the $A_1$ and $A_2$ irrep labels of the flat bands at the zone center are interchanged \cite{Song19,Hejazi19}. This is similar to the photocurrent direction change in a semiconductor or insulator when the conduction and valence bands are inverted ~\cite{Tan16,Cook17,Yan18,Sivianes23}, but notably also occurs in an approximate way when the bands inverted are both occupied and optical transitions occur to a third band.  
This suggests that these optical responses can be used to determine whether a given sample is above or below the magic angle in an absolute way, assuming the sign of the twist angle is known (the overall sign of $\sigma_{xyz}$ and $\eta_{xyz}$ flips for negative $\theta$). Alternatively, the effect can be used to screen for samples with the angle closest to the magic one by minimizing either response. 

\emph{Extrinsic photocurrents} - The presence of a substrate or encapsulation gives rise to the three new components of the photocurrent tensors listed in Table \ref{components}, which we now analyze. Figure \ref{Fig3} shows all three components in the presence of the corresponding perturbation that generates it, as a function of the chemical potential, for two representative values of $\theta$ above and below the magic angle, and three values of the chemical potential corresponding to neutrality ($\mu=0$), filled flat bands ($\mu>0$), and empty flat bands ($\mu<0$). Consistent with previous literature \cite{Kaplan22,Chaudhary22}, we find extremely large normal incidence photocurrents reaching $10^5$ ${\rm nm \; \mu A/V^2}$, especially when both flat bands are occupied, see Figs. \ref{Fig3} a,b. Off-normal incidence photocurrents induced by an \red{electric} field are in contrast smaller and comparable to the intrinsic components. We observe that the particle-hole constraints are obeyed in every case, with only $\sigma_{yyy}$ inverting its sign with $\mu$. Interestingly, we also observe that $\sigma_{xxx}$ and $\sigma_{yyy}$ display an approximate reversal accross the magic angle, while $\sigma_{xxz}$ and $\eta_{xxz}$ do not. 

\emph{Breaking particle-hole symmetry} - The continuum Hamiltonian in  Eq. \ref{continuummodel}, derived in the small $\theta$ approximation, has an exact PHS. This symmetry is weakly broken once subleading corrections in $\theta$ are considered, and we now assess their effect.  

The first of such corrections is the relative rotation of the spinor basis on each layer, so that at small angles
\begin{align}
\boldsymbol{\sigma}_{\nu}\cdot\boldsymbol{\partial} \rightarrow \boldsymbol{\sigma}_{\nu}\cdot\boldsymbol{\partial}+\frac{\theta}{2}\left(\boldsymbol{\sigma}_{\nu}\times\boldsymbol{\partial}\right)_z\tau_z.
\label{paulirot}
\end{align}
The second correction is an interlayer tunneling term between the same sublattices introduced in Refs. \cite{Vafek20RG,KangVafekPH23} by Kang and Vafek,
\begin{align}
    T_{n, \nu} \rightarrow T_{n, \nu} + i w_3  \nu \sigma_z. \label{kangvafek}
\end{align}
\red{This term is expected to arise due to lattice relaxation.}

It was shown in Ref.~\onlinecite{Scheer22} that these two terms can be transformed into each other by a unitary transformation $e^{i\sigma_z \tau_z \alpha}$, which does not affect the computation of the photocurrents. Therefore, we just include Eq. \ref{kangvafek} with value $w_3=0.9$ meV \cite{KangVafekPH23} as a representative example of a realistic band structure with weak PHS breaking. PHS can also be broken due to more complex corrections \red{introduced by lattice relaxation} like non local ($k$-dependent) interlayer tunneling~\cite{Guinea19,Fang19,Koshino20}, with a similar effect on the bands \cite{KangVafekPH23}, which we do not consider for simplicity. 

\begin{figure}[t]
    \centering
\includegraphics[width=0.45\textwidth]{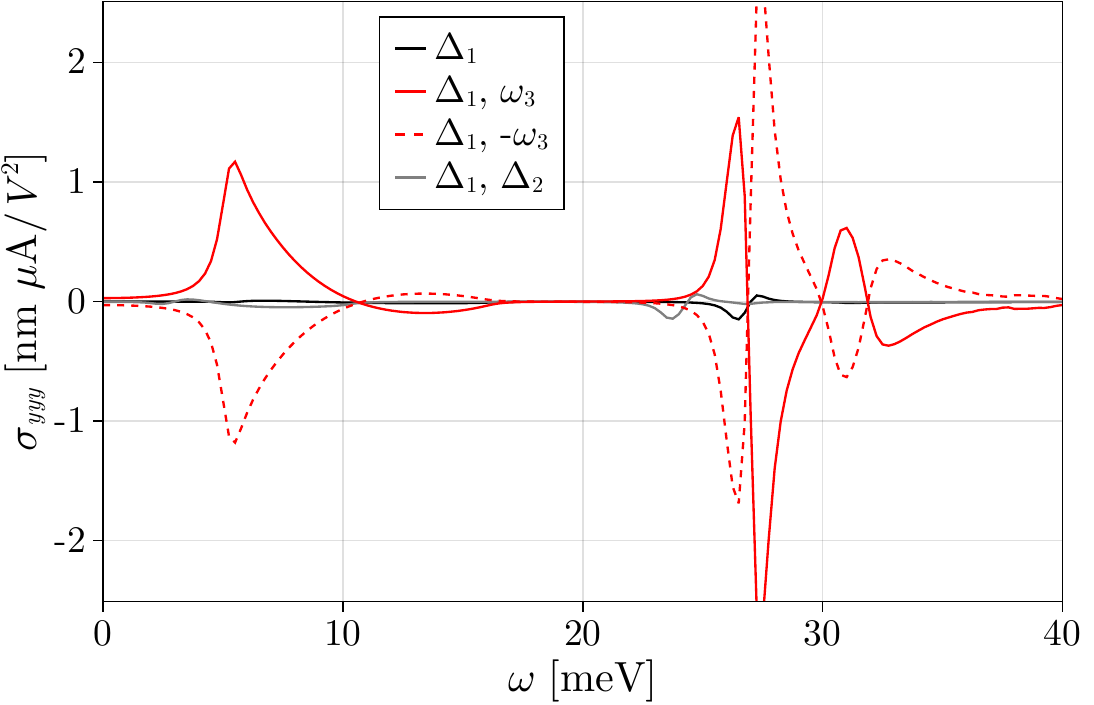}
     \caption{$\sigma_{yyy}$ in the presence of PHS breaking perturbations at neutrality. A $\Delta_1\sigma_z$ mass in all simulations allows for a finite extrinsic response. Two types of PHS breaking terms are considered: a layered-resolved sublattice perturbation, $\Delta_1 \sigma_z + \Delta_2 \sigma_z\tau_z$, with $\Delta_1 = -\Delta_2 = 10$ meV (gray), and the Kang-Vafek term in Eq. \eqref{kangvafek} (red). Solid and dashed red lines correspond to $\omega_3 = \pm0.9$ meV, respectively. The PHS preserving case is displayed in black as reference. We set $\theta = 1.05^\circ$.}
    \label{Fig4}
\end{figure}

Equation \ref{PHSyyy} shows that $\sigma_{yyy}$ is the only component that flips sign with $\mu$ due to PHS and should vanish at neutrality (in the presence of $\Delta_1$), and is therefore the ideal component to assess the effect of PHS breaking. Figure \ref{Fig4} considers the change in this component in the presence of finite $w_3$ at neutrality. The characteristic spectrum found serves as a proxy to track $w_3$, with a response that is an order of magnitude less than $\sigma_{yyy}$ at finite $\mu$, but still quite large to be detected. In Ref.~\onlinecite{SM} we also show how the rest of components are only weakly corrected by the presence of finite $w_3$.

\emph{Discussion} - In this work we have introduced the intrinsic photocurrents of TBG, showing they are of sizable magnitude and that they can be used to track the proximity to the magic angle. It is interesting to compare our results with the predictions for circular dichroism (CD)~\cite{Morell17,Stauber18},
where a reversal of the intraband contribution is also observed when crossing the magic angle~\cite{Stauber20}. The key difference between the intrinsic photocurrents and CD is that CD vanishes with PHS~\cite{Morell17}, so its sign is determined both by the chemical potential and weak PHS breaking perturbations, and it cannot be used to determine the sign of $\theta-\theta_M$ in an absolute way. The intrinsic photocurrents do not change sign with chemical potential or weak PHS breaking, and thus track the sign of $\theta-\theta_M$ directly.

These intrisic photocurrents can also serve as a benchmark to gauge the amount of symmetry breaking by comparing the orders of magnitude of extrinsic vs. intrinsic photocurrents. We have also shown how $\sigma_{yyy}$ can be used as a good proxy of PHS  breaking. Away from neutrality PHS breaking can be enhanced by Hartree renormalization, which can lead to changes in the photocurrents \cite{Chaudhary22} which we anticipate can also be tracked by $\sigma_{yyy}$. We also expect that at low temperatures, our partitioning of extrinsic and intrisic currents can also help diagnose more complex symmetry breaking patterns induced by interactions. The computation of such interacting photocurrents however represents an important challenge, as effective models where interactions are more tractable than in the continuum model do not often keep track of the layer degree of freedom, or do not implement PH symmetry. In summary, we expect the comparison of intrinsic vs extrinsic photocurrents to become a versatile tool to characterize twist angles and symmetry breaking patterns in TBG.

\emph{Data Availability} - Computer codes, raw data and analysis scripts for all presented figures are available in the Zenodo database under accession code: https://zenodo.org/records/17702853

\emph{Acknowledgements} - We thank O. Pozo, O. Vafek, D. Kaplan, T. Holder for insightful discussions. This work is supported by Grant PID2021-128760NB0-I00 from the Spanish MCIN/AEI/10.13039/501100011033/FEDER, EU. 

\begin{thebibliography}{55}%
\makeatletter
\providecommand \@ifxundefined [1]{%
 \@ifx{#1\undefined}
}%
\providecommand \@ifnum [1]{%
 \ifnum #1\expandafter \@firstoftwo
 \else \expandafter \@secondoftwo
 \fi
}%
\providecommand \@ifx [1]{%
 \ifx #1\expandafter \@firstoftwo
 \else \expandafter \@secondoftwo
 \fi
}%
\providecommand \natexlab [1]{#1}%
\providecommand \enquote  [1]{``#1''}%
\providecommand \bibnamefont  [1]{#1}%
\providecommand \bibfnamefont [1]{#1}%
\providecommand \citenamefont [1]{#1}%
\providecommand \href@noop [0]{\@secondoftwo}%
\providecommand \href [0]{\begingroup \@sanitize@url \@href}%
\providecommand \@href[1]{\@@startlink{#1}\@@href}%
\providecommand \@@href[1]{\endgroup#1\@@endlink}%
\providecommand \@sanitize@url [0]{\catcode `\\12\catcode `\$12\catcode
  `\&12\catcode `\#12\catcode `\^12\catcode `\_12\catcode `\%12\relax}%
\providecommand \@@startlink[1]{}%
\providecommand \@@endlink[0]{}%
\providecommand \url  [0]{\begingroup\@sanitize@url \@url }%
\providecommand \@url [1]{\endgroup\@href {#1}{\urlprefix }}%
\providecommand \urlprefix  [0]{URL }%
\providecommand \Eprint [0]{\href }%
\providecommand \doibase [0]{https://doi.org/}%
\providecommand \selectlanguage [0]{\@gobble}%
\providecommand \bibinfo  [0]{\@secondoftwo}%
\providecommand \bibfield  [0]{\@secondoftwo}%
\providecommand \translation [1]{[#1]}%
\providecommand \BibitemOpen [0]{}%
\providecommand \bibitemStop [0]{}%
\providecommand \bibitemNoStop [0]{.\EOS\space}%
\providecommand \EOS [0]{\spacefactor3000\relax}%
\providecommand \BibitemShut  [1]{\csname bibitem#1\endcsname}%
\let\auto@bib@innerbib\@empty
\bibitem [{\citenamefont {Bistritzer}\ and\ \citenamefont
  {MacDonald}(2011)}]{Bistritzer11}%
  \BibitemOpen
  \bibfield  {author} {\bibinfo {author} {\bibfnamefont {R.}~\bibnamefont
  {Bistritzer}}\ and\ \bibinfo {author} {\bibfnamefont {A.~H.}\ \bibnamefont
  {MacDonald}},\ }\bibfield  {title} {\bibinfo {title} {Moir{\'e} bands in
  twisted double-layer graphene},\ }\href
  {https://doi.org/10.1073/pnas.1108174108} {\bibfield  {journal} {\bibinfo
  {journal} {Proceedings of the National Academy of Sciences}\ }\textbf
  {\bibinfo {volume} {108}},\ \bibinfo {pages} {12233} (\bibinfo {year}
  {2011})}\BibitemShut {NoStop}%
\bibitem [{\citenamefont {Cao}\ \emph {et~al.}(2018{\natexlab{a}})\citenamefont
  {Cao}, \citenamefont {Fatemi}, \citenamefont {Demir}, \citenamefont {Fang},
  \citenamefont {Tomarken}, \citenamefont {Luo}, \citenamefont
  {Sanchez-Yamagishi}, \citenamefont {Watanabe}, \citenamefont {Taniguchi},
  \citenamefont {Kaxiras} \emph {et~al.}}]{Cao2018Insulators}%
  \BibitemOpen
  \bibfield  {author} {\bibinfo {author} {\bibfnamefont {Y.}~\bibnamefont
  {Cao}}, \bibinfo {author} {\bibfnamefont {V.}~\bibnamefont {Fatemi}},
  \bibinfo {author} {\bibfnamefont {A.}~\bibnamefont {Demir}}, \bibinfo
  {author} {\bibfnamefont {S.}~\bibnamefont {Fang}}, \bibinfo {author}
  {\bibfnamefont {S.~L.}\ \bibnamefont {Tomarken}}, \bibinfo {author}
  {\bibfnamefont {J.~Y.}\ \bibnamefont {Luo}}, \bibinfo {author} {\bibfnamefont
  {J.~D.}\ \bibnamefont {Sanchez-Yamagishi}}, \bibinfo {author} {\bibfnamefont
  {K.}~\bibnamefont {Watanabe}}, \bibinfo {author} {\bibfnamefont
  {T.}~\bibnamefont {Taniguchi}}, \bibinfo {author} {\bibfnamefont
  {E.}~\bibnamefont {Kaxiras}}, \emph {et~al.},\ }\bibfield  {title} {\bibinfo
  {title} {Correlated insulator behaviour at half-filling in magic-angle
  graphene superlattices},\ }\href@noop {} {\bibfield  {journal} {\bibinfo
  {journal} {Nature}\ }\textbf {\bibinfo {volume} {556}},\ \bibinfo {pages}
  {80} (\bibinfo {year} {2018}{\natexlab{a}})}\BibitemShut {NoStop}%
\bibitem [{\citenamefont {Cao}\ \emph {et~al.}(2018{\natexlab{b}})\citenamefont
  {Cao}, \citenamefont {Fatemi}, \citenamefont {Fang}, \citenamefont
  {Watanabe}, \citenamefont {Taniguchi}, \citenamefont {Kaxiras},\ and\
  \citenamefont {Jarillo-Herrero}}]{Cao18SC}%
  \BibitemOpen
  \bibfield  {author} {\bibinfo {author} {\bibfnamefont {Y.}~\bibnamefont
  {Cao}}, \bibinfo {author} {\bibfnamefont {V.}~\bibnamefont {Fatemi}},
  \bibinfo {author} {\bibfnamefont {S.}~\bibnamefont {Fang}}, \bibinfo {author}
  {\bibfnamefont {K.}~\bibnamefont {Watanabe}}, \bibinfo {author}
  {\bibfnamefont {T.}~\bibnamefont {Taniguchi}}, \bibinfo {author}
  {\bibfnamefont {E.}~\bibnamefont {Kaxiras}},\ and\ \bibinfo {author}
  {\bibfnamefont {P.}~\bibnamefont {Jarillo-Herrero}},\ }\bibfield  {title}
  {\bibinfo {title} {Unconventional superconductivity in magic-angle graphene
  superlattices},\ }\href@noop {} {\bibfield  {journal} {\bibinfo  {journal}
  {Nature}\ }\textbf {\bibinfo {volume} {556}},\ \bibinfo {pages} {43}
  (\bibinfo {year} {2018}{\natexlab{b}})}\BibitemShut {NoStop}%
\bibitem [{\citenamefont {Kim}\ \emph {et~al.}(2016)\citenamefont {Kim},
  \citenamefont {S{\'a}nchez-Castillo}, \citenamefont {Ziegler}, \citenamefont
  {Ogawa}, \citenamefont {Noguez},\ and\ \citenamefont {Park}}]{Kim16}%
  \BibitemOpen
  \bibfield  {author} {\bibinfo {author} {\bibfnamefont {C.-J.}\ \bibnamefont
  {Kim}}, \bibinfo {author} {\bibfnamefont {A.}~\bibnamefont
  {S{\'a}nchez-Castillo}}, \bibinfo {author} {\bibfnamefont {Z.}~\bibnamefont
  {Ziegler}}, \bibinfo {author} {\bibfnamefont {Y.}~\bibnamefont {Ogawa}},
  \bibinfo {author} {\bibfnamefont {C.}~\bibnamefont {Noguez}},\ and\ \bibinfo
  {author} {\bibfnamefont {J.}~\bibnamefont {Park}},\ }\bibfield  {title}
  {\bibinfo {title} {Chiral atomically thin films},\ }\href@noop {} {\bibfield
  {journal} {\bibinfo  {journal} {Nature nanotechnology}\ }\textbf {\bibinfo
  {volume} {11}},\ \bibinfo {pages} {520} (\bibinfo {year} {2016})}\BibitemShut
  {NoStop}%
\bibitem [{\citenamefont {Liu}\ and\ \citenamefont
  {Dai}(2020)}]{Liu20Anomalous}%
  \BibitemOpen
  \bibfield  {author} {\bibinfo {author} {\bibfnamefont {J.}~\bibnamefont
  {Liu}}\ and\ \bibinfo {author} {\bibfnamefont {X.}~\bibnamefont {Dai}},\
  }\bibfield  {title} {\bibinfo {title} {Anomalous hall effect, magneto-optical
  properties, and nonlinear optical properties of twisted graphene systems},\
  }\href@noop {} {\bibfield  {journal} {\bibinfo  {journal} {npj Comp. Mater.}\
  }\textbf {\bibinfo {volume} {6}},\ \bibinfo {pages} {57} (\bibinfo {year}
  {2020})}\BibitemShut {NoStop}%
\bibitem [{\citenamefont {Kaplan}\ \emph {et~al.}(2022)\citenamefont {Kaplan},
  \citenamefont {Holder},\ and\ \citenamefont {Yan}}]{Kaplan22}%
  \BibitemOpen
  \bibfield  {author} {\bibinfo {author} {\bibfnamefont {D.}~\bibnamefont
  {Kaplan}}, \bibinfo {author} {\bibfnamefont {T.}~\bibnamefont {Holder}},\
  and\ \bibinfo {author} {\bibfnamefont {B.}~\bibnamefont {Yan}},\ }\bibfield
  {title} {\bibinfo {title} {Twisted photovoltaics at terahertz frequencies
  from momentum shift current},\ }\href
  {https://doi.org/10.1103/PhysRevResearch.4.013209} {\bibfield  {journal}
  {\bibinfo  {journal} {Phys. Rev. Res.}\ }\textbf {\bibinfo {volume} {4}},\
  \bibinfo {pages} {013209} (\bibinfo {year} {2022})}\BibitemShut {NoStop}%
\bibitem [{\citenamefont {Chaudhary}\ \emph {et~al.}(2022)\citenamefont
  {Chaudhary}, \citenamefont {Lewandowski},\ and\ \citenamefont
  {Refael}}]{Chaudhary22}%
  \BibitemOpen
  \bibfield  {author} {\bibinfo {author} {\bibfnamefont {S.}~\bibnamefont
  {Chaudhary}}, \bibinfo {author} {\bibfnamefont {C.}~\bibnamefont
  {Lewandowski}},\ and\ \bibinfo {author} {\bibfnamefont {G.}~\bibnamefont
  {Refael}},\ }\bibfield  {title} {\bibinfo {title} {Shift-current response as
  a probe of quantum geometry and electron-electron interactions in twisted
  bilayer graphene},\ }\href {https://doi.org/10.1103/PhysRevResearch.4.013164}
  {\bibfield  {journal} {\bibinfo  {journal} {Phys. Rev. Res.}\ }\textbf
  {\bibinfo {volume} {4}},\ \bibinfo {pages} {013164} (\bibinfo {year}
  {2022})}\BibitemShut {NoStop}%
\bibitem [{\citenamefont {Zhang}\ \emph {et~al.}(2022)\citenamefont {Zhang},
  \citenamefont {Lu},\ and\ \citenamefont {Liu}}]{Zhang22Correlated}%
  \BibitemOpen
  \bibfield  {author} {\bibinfo {author} {\bibfnamefont {S.}~\bibnamefont
  {Zhang}}, \bibinfo {author} {\bibfnamefont {X.}~\bibnamefont {Lu}},\ and\
  \bibinfo {author} {\bibfnamefont {J.}~\bibnamefont {Liu}},\ }\bibfield
  {title} {\bibinfo {title} {Correlated insulators, density wave states, and
  their nonlinear optical response in magic-angle twisted bilayer graphene},\
  }\href {https://doi.org/10.1103/PhysRevLett.128.247402} {\bibfield  {journal}
  {\bibinfo  {journal} {Phys. Rev. Lett.}\ }\textbf {\bibinfo {volume} {128}},\
  \bibinfo {pages} {247402} (\bibinfo {year} {2022})}\BibitemShut {NoStop}%
\bibitem [{\citenamefont {Yang}\ \emph {et~al.}(2020)\citenamefont {Yang},
  \citenamefont {Song}, \citenamefont {Meng}, \citenamefont {Luo},
  \citenamefont {Lou}, \citenamefont {Lin}, \citenamefont {Gong}, \citenamefont
  {Cao}, \citenamefont {Barnard}, \citenamefont {Chan} \emph
  {et~al.}}]{Yang20}%
  \BibitemOpen
  \bibfield  {author} {\bibinfo {author} {\bibfnamefont {F.}~\bibnamefont
  {Yang}}, \bibinfo {author} {\bibfnamefont {W.}~\bibnamefont {Song}}, \bibinfo
  {author} {\bibfnamefont {F.}~\bibnamefont {Meng}}, \bibinfo {author}
  {\bibfnamefont {F.}~\bibnamefont {Luo}}, \bibinfo {author} {\bibfnamefont
  {S.}~\bibnamefont {Lou}}, \bibinfo {author} {\bibfnamefont {S.}~\bibnamefont
  {Lin}}, \bibinfo {author} {\bibfnamefont {Z.}~\bibnamefont {Gong}}, \bibinfo
  {author} {\bibfnamefont {J.}~\bibnamefont {Cao}}, \bibinfo {author}
  {\bibfnamefont {E.~S.}\ \bibnamefont {Barnard}}, \bibinfo {author}
  {\bibfnamefont {E.}~\bibnamefont {Chan}}, \emph {et~al.},\ }\bibfield
  {title} {\bibinfo {title} {Tunable second harmonic generation in twisted
  bilayer graphene},\ }\href@noop {} {\bibfield  {journal} {\bibinfo  {journal}
  {Matter}\ }\textbf {\bibinfo {volume} {3}},\ \bibinfo {pages} {1361}
  (\bibinfo {year} {2020})}\BibitemShut {NoStop}%
\bibitem [{\citenamefont {Morell}\ \emph {et~al.}(2017)\citenamefont {Morell},
  \citenamefont {Chico},\ and\ \citenamefont {Brey}}]{Morell17}%
  \BibitemOpen
  \bibfield  {author} {\bibinfo {author} {\bibfnamefont {E.~S.}\ \bibnamefont
  {Morell}}, \bibinfo {author} {\bibfnamefont {L.}~\bibnamefont {Chico}},\ and\
  \bibinfo {author} {\bibfnamefont {L.}~\bibnamefont {Brey}},\ }\bibfield
  {title} {\bibinfo {title} {Twisting dirac fermions: circular dichroism in
  bilayer graphene},\ }\href@noop {} {\bibfield  {journal} {\bibinfo  {journal}
  {2D Materials}\ }\textbf {\bibinfo {volume} {4}},\ \bibinfo {pages} {035015}
  (\bibinfo {year} {2017})}\BibitemShut {NoStop}%
\bibitem [{\citenamefont {Stauber}\ \emph {et~al.}(2018)\citenamefont
  {Stauber}, \citenamefont {Low},\ and\ \citenamefont
  {G\'omez-Santos}}]{Stauber18}%
  \BibitemOpen
  \bibfield  {author} {\bibinfo {author} {\bibfnamefont {T.}~\bibnamefont
  {Stauber}}, \bibinfo {author} {\bibfnamefont {T.}~\bibnamefont {Low}},\ and\
  \bibinfo {author} {\bibfnamefont {G.}~\bibnamefont {G\'omez-Santos}},\
  }\bibfield  {title} {\bibinfo {title} {Chiral response of twisted bilayer
  graphene},\ }\href {https://doi.org/10.1103/PhysRevLett.120.046801}
  {\bibfield  {journal} {\bibinfo  {journal} {Phys. Rev. Lett.}\ }\textbf
  {\bibinfo {volume} {120}},\ \bibinfo {pages} {046801} (\bibinfo {year}
  {2018})}\BibitemShut {NoStop}%
\bibitem [{\citenamefont {Stauber}\ \emph {et~al.}(2020)\citenamefont
  {Stauber}, \citenamefont {Gonz\'alez},\ and\ \citenamefont
  {G\'omez-Santos}}]{Stauber20}%
  \BibitemOpen
  \bibfield  {author} {\bibinfo {author} {\bibfnamefont {T.}~\bibnamefont
  {Stauber}}, \bibinfo {author} {\bibfnamefont {J.}~\bibnamefont
  {Gonz\'alez}},\ and\ \bibinfo {author} {\bibfnamefont {G.}~\bibnamefont
  {G\'omez-Santos}},\ }\bibfield  {title} {\bibinfo {title} {Change of
  chirality at magic angles of twisted bilayer graphene},\ }\href
  {https://doi.org/10.1103/PhysRevB.102.081404} {\bibfield  {journal} {\bibinfo
   {journal} {Phys. Rev. B}\ }\textbf {\bibinfo {volume} {102}},\ \bibinfo
  {pages} {081404(R)} (\bibinfo {year} {2020})}\BibitemShut {NoStop}%
\bibitem [{\citenamefont {Wang}\ \emph {et~al.}(2020)\citenamefont {Wang},
  \citenamefont {Morimoto},\ and\ \citenamefont {Moore}}]{Wang20Optical}%
  \BibitemOpen
  \bibfield  {author} {\bibinfo {author} {\bibfnamefont {Y.-Q.}\ \bibnamefont
  {Wang}}, \bibinfo {author} {\bibfnamefont {T.}~\bibnamefont {Morimoto}},\
  and\ \bibinfo {author} {\bibfnamefont {J.~E.}\ \bibnamefont {Moore}},\
  }\bibfield  {title} {\bibinfo {title} {Optical rotation in thin
  chiral/twisted materials and the gyrotropic magnetic effect},\ }\href
  {https://doi.org/10.1103/PhysRevB.101.174419} {\bibfield  {journal} {\bibinfo
   {journal} {Phys. Rev. B}\ }\textbf {\bibinfo {volume} {101}},\ \bibinfo
  {pages} {174419} (\bibinfo {year} {2020})}\BibitemShut {NoStop}%
\bibitem [{\citenamefont {Chang}\ \emph {et~al.}(2022)\citenamefont {Chang},
  \citenamefont {Zheng}, \citenamefont {Sipe},\ and\ \citenamefont
  {Cheng}}]{Chang22}%
  \BibitemOpen
  \bibfield  {author} {\bibinfo {author} {\bibfnamefont {K.}~\bibnamefont
  {Chang}}, \bibinfo {author} {\bibfnamefont {Z.}~\bibnamefont {Zheng}},
  \bibinfo {author} {\bibfnamefont {J.~E.}\ \bibnamefont {Sipe}},\ and\
  \bibinfo {author} {\bibfnamefont {J.~L.}\ \bibnamefont {Cheng}},\ }\bibfield
  {title} {\bibinfo {title} {Theory of optical activity in doped systems with
  application to twisted bilayer graphene},\ }\href
  {https://doi.org/10.1103/PhysRevB.106.245405} {\bibfield  {journal} {\bibinfo
   {journal} {Phys. Rev. B}\ }\textbf {\bibinfo {volume} {106}},\ \bibinfo
  {pages} {245405} (\bibinfo {year} {2022})}\BibitemShut {NoStop}%
\bibitem [{\citenamefont {Yuan}\ \emph {et~al.}(2014)\citenamefont {Yuan},
  \citenamefont {Wang}, \citenamefont {Lian}, \citenamefont {Zhang},
  \citenamefont {Fang}, \citenamefont {Shen}, \citenamefont {Xu}, \citenamefont
  {Xu}, \citenamefont {Zhang}, \citenamefont {Hwang} \emph {et~al.}}]{Yuan14}%
  \BibitemOpen
  \bibfield  {author} {\bibinfo {author} {\bibfnamefont {H.}~\bibnamefont
  {Yuan}}, \bibinfo {author} {\bibfnamefont {X.}~\bibnamefont {Wang}}, \bibinfo
  {author} {\bibfnamefont {B.}~\bibnamefont {Lian}}, \bibinfo {author}
  {\bibfnamefont {H.}~\bibnamefont {Zhang}}, \bibinfo {author} {\bibfnamefont
  {X.}~\bibnamefont {Fang}}, \bibinfo {author} {\bibfnamefont {B.}~\bibnamefont
  {Shen}}, \bibinfo {author} {\bibfnamefont {G.}~\bibnamefont {Xu}}, \bibinfo
  {author} {\bibfnamefont {Y.}~\bibnamefont {Xu}}, \bibinfo {author}
  {\bibfnamefont {S.-C.}\ \bibnamefont {Zhang}}, \bibinfo {author}
  {\bibfnamefont {H.~Y.}\ \bibnamefont {Hwang}}, \emph {et~al.},\ }\bibfield
  {title} {\bibinfo {title} {Generation and electric control of
  spin--valley-coupled circular photogalvanic current in wse2},\ }\href@noop {}
  {\bibfield  {journal} {\bibinfo  {journal} {Nat. Nanotech.}\ }\textbf
  {\bibinfo {volume} {9}},\ \bibinfo {pages} {851} (\bibinfo {year}
  {2014})}\BibitemShut {NoStop}%
\bibitem [{\citenamefont {Quereda}\ \emph {et~al.}(2018)\citenamefont
  {Quereda}, \citenamefont {Ghiasi}, \citenamefont {You}, \citenamefont
  {van~den Brink}, \citenamefont {van Wees},\ and\ \citenamefont {van~der
  Wal}}]{Quereda18}%
  \BibitemOpen
  \bibfield  {author} {\bibinfo {author} {\bibfnamefont {J.}~\bibnamefont
  {Quereda}}, \bibinfo {author} {\bibfnamefont {T.~S.}\ \bibnamefont {Ghiasi}},
  \bibinfo {author} {\bibfnamefont {J.-S.}\ \bibnamefont {You}}, \bibinfo
  {author} {\bibfnamefont {J.}~\bibnamefont {van~den Brink}}, \bibinfo {author}
  {\bibfnamefont {B.~J.}\ \bibnamefont {van Wees}},\ and\ \bibinfo {author}
  {\bibfnamefont {C.~H.}\ \bibnamefont {van~der Wal}},\ }\bibfield  {title}
  {\bibinfo {title} {Symmetry regimes for circular photocurrents in monolayer
  mose2},\ }\href@noop {} {\bibfield  {journal} {\bibinfo  {journal} {Nat.
  Commun.}\ }\textbf {\bibinfo {volume} {9}},\ \bibinfo {pages} {3346}
  (\bibinfo {year} {2018})}\BibitemShut {NoStop}%
\bibitem [{\citenamefont {Ni}\ \emph {et~al.}(2021)\citenamefont {Ni},
  \citenamefont {Wang}, \citenamefont {Zhang}, \citenamefont {Pozo},
  \citenamefont {Xu}, \citenamefont {Han}, \citenamefont {Manna}, \citenamefont
  {Paglione}, \citenamefont {Felser}, \citenamefont {Grushin} \emph
  {et~al.}}]{Ni21}%
  \BibitemOpen
  \bibfield  {author} {\bibinfo {author} {\bibfnamefont {Z.}~\bibnamefont
  {Ni}}, \bibinfo {author} {\bibfnamefont {K.}~\bibnamefont {Wang}}, \bibinfo
  {author} {\bibfnamefont {Y.}~\bibnamefont {Zhang}}, \bibinfo {author}
  {\bibfnamefont {O.}~\bibnamefont {Pozo}}, \bibinfo {author} {\bibfnamefont
  {B.}~\bibnamefont {Xu}}, \bibinfo {author} {\bibfnamefont {X.}~\bibnamefont
  {Han}}, \bibinfo {author} {\bibfnamefont {K.}~\bibnamefont {Manna}}, \bibinfo
  {author} {\bibfnamefont {J.}~\bibnamefont {Paglione}}, \bibinfo {author}
  {\bibfnamefont {C.}~\bibnamefont {Felser}}, \bibinfo {author} {\bibfnamefont
  {A.~G.}\ \bibnamefont {Grushin}}, \emph {et~al.},\ }\bibfield  {title}
  {\bibinfo {title} {Giant topological longitudinal circular photo-galvanic
  effect in the chiral multifold semimetal cosi},\ }\href@noop {} {\bibfield
  {journal} {\bibinfo  {journal} {Nat. Commun.}\ }\textbf {\bibinfo {volume}
  {12}},\ \bibinfo {pages} {154} (\bibinfo {year} {2021})}\BibitemShut
  {NoStop}%
\bibitem [{\citenamefont {Chen}\ \emph {et~al.}(2023)\citenamefont {Chen},
  \citenamefont {Zhai}, \citenamefont {Xiao},\ and\ \citenamefont
  {Yao}}]{Chen23Crossed}%
  \BibitemOpen
  \bibfield  {author} {\bibinfo {author} {\bibfnamefont {C.}~\bibnamefont
  {Chen}}, \bibinfo {author} {\bibfnamefont {D.}~\bibnamefont {Zhai}}, \bibinfo
  {author} {\bibfnamefont {C.}~\bibnamefont {Xiao}},\ and\ \bibinfo {author}
  {\bibfnamefont {W.}~\bibnamefont {Yao}},\ }\bibfield  {title} {\bibinfo
  {title} {Crossed nonlinear dynamical hall effect in twisted bilayers},\
  }\href@noop {} {\bibfield  {journal} {\bibinfo  {journal} {arXiv preprint
  arXiv:2303.09973}\ } (\bibinfo {year} {2023})}\BibitemShut {NoStop}%
\bibitem [{\citenamefont {Zheng}\ \emph {et~al.}(2023)\citenamefont {Zheng},
  \citenamefont {Chang},\ and\ \citenamefont {Cheng}}]{Zheng2023}%
  \BibitemOpen
  \bibfield  {author} {\bibinfo {author} {\bibfnamefont {Z.}~\bibnamefont
  {Zheng}}, \bibinfo {author} {\bibfnamefont {K.}~\bibnamefont {Chang}},\ and\
  \bibinfo {author} {\bibfnamefont {J.~L.}\ \bibnamefont {Cheng}},\ }\href@noop
  {} {\bibinfo {title} {Gate voltage induced injection and shift currents in
  aa- and ab-stacked bilayer graphene}} (\bibinfo {year} {2023})\BibitemShut
  {NoStop}%
\bibitem [{\citenamefont {Arora}\ \emph {et~al.}(2021)\citenamefont {Arora},
  \citenamefont {Kong},\ and\ \citenamefont {Song}}]{Arora21}%
  \BibitemOpen
  \bibfield  {author} {\bibinfo {author} {\bibfnamefont {A.}~\bibnamefont
  {Arora}}, \bibinfo {author} {\bibfnamefont {J.~F.}\ \bibnamefont {Kong}},\
  and\ \bibinfo {author} {\bibfnamefont {J.~C.~W.}\ \bibnamefont {Song}},\
  }\bibfield  {title} {\bibinfo {title} {Strain-induced large injection current
  in twisted bilayer graphene},\ }\href
  {https://doi.org/10.1103/PhysRevB.104.L241404} {\bibfield  {journal}
  {\bibinfo  {journal} {Phys. Rev. B}\ }\textbf {\bibinfo {volume} {104}},\
  \bibinfo {pages} {L241404} (\bibinfo {year} {2021})}\BibitemShut {NoStop}%
\bibitem [{\citenamefont {Hesp}\ \emph {et~al.}(2021)\citenamefont {Hesp},
  \citenamefont {Torre}, \citenamefont {Barcons-Ruiz}, \citenamefont
  {Herzig~Sheinfux}, \citenamefont {Watanabe}, \citenamefont {Taniguchi},
  \citenamefont {Krishna~Kumar},\ and\ \citenamefont {Koppens}}]{Hesp21}%
  \BibitemOpen
  \bibfield  {author} {\bibinfo {author} {\bibfnamefont {N.~C.}\ \bibnamefont
  {Hesp}}, \bibinfo {author} {\bibfnamefont {I.}~\bibnamefont {Torre}},
  \bibinfo {author} {\bibfnamefont {D.}~\bibnamefont {Barcons-Ruiz}}, \bibinfo
  {author} {\bibfnamefont {H.}~\bibnamefont {Herzig~Sheinfux}}, \bibinfo
  {author} {\bibfnamefont {K.}~\bibnamefont {Watanabe}}, \bibinfo {author}
  {\bibfnamefont {T.}~\bibnamefont {Taniguchi}}, \bibinfo {author}
  {\bibfnamefont {R.}~\bibnamefont {Krishna~Kumar}},\ and\ \bibinfo {author}
  {\bibfnamefont {F.~H.}\ \bibnamefont {Koppens}},\ }\bibfield  {title}
  {\bibinfo {title} {Nano-imaging photoresponse in a moir{\'e} unit cell of
  minimally twisted bilayer graphene},\ }\href@noop {} {\bibfield  {journal}
  {\bibinfo  {journal} {Nat. Commun.}\ }\textbf {\bibinfo {volume} {12}},\
  \bibinfo {pages} {1640} (\bibinfo {year} {2021})}\BibitemShut {NoStop}%
\bibitem [{\citenamefont {Sunku}\ \emph {et~al.}(2021)\citenamefont {Sunku},
  \citenamefont {Halbertal}, \citenamefont {Stauber}, \citenamefont {Chen},
  \citenamefont {McLeod}, \citenamefont {Rikhter}, \citenamefont {Berkowitz},
  \citenamefont {Lo}, \citenamefont {Gonzalez-Acevedo}, \citenamefont {Hone}
  \emph {et~al.}}]{Sunku21}%
  \BibitemOpen
  \bibfield  {author} {\bibinfo {author} {\bibfnamefont {S.~S.}\ \bibnamefont
  {Sunku}}, \bibinfo {author} {\bibfnamefont {D.}~\bibnamefont {Halbertal}},
  \bibinfo {author} {\bibfnamefont {T.}~\bibnamefont {Stauber}}, \bibinfo
  {author} {\bibfnamefont {S.}~\bibnamefont {Chen}}, \bibinfo {author}
  {\bibfnamefont {A.~S.}\ \bibnamefont {McLeod}}, \bibinfo {author}
  {\bibfnamefont {A.}~\bibnamefont {Rikhter}}, \bibinfo {author} {\bibfnamefont
  {M.~E.}\ \bibnamefont {Berkowitz}}, \bibinfo {author} {\bibfnamefont
  {C.~F.~B.}\ \bibnamefont {Lo}}, \bibinfo {author} {\bibfnamefont {D.~E.}\
  \bibnamefont {Gonzalez-Acevedo}}, \bibinfo {author} {\bibfnamefont {J.~C.}\
  \bibnamefont {Hone}}, \emph {et~al.},\ }\bibfield  {title} {\bibinfo {title}
  {Hyperbolic enhancement of photocurrent patterns in minimally twisted bilayer
  graphene},\ }\href@noop {} {\bibfield  {journal} {\bibinfo  {journal} {Nat.
  Commun.}\ }\textbf {\bibinfo {volume} {12}},\ \bibinfo {pages} {1641}
  (\bibinfo {year} {2021})}\BibitemShut {NoStop}%
\bibitem [{\citenamefont {Otteneder}\ \emph {et~al.}(2020)\citenamefont
  {Otteneder}, \citenamefont {Hubmann}, \citenamefont {Lu}, \citenamefont
  {Kozlov}, \citenamefont {Golub}, \citenamefont {Watanabe}, \citenamefont
  {Taniguchi}, \citenamefont {Efetov},\ and\ \citenamefont
  {Ganichev}}]{Otteneder20}%
  \BibitemOpen
  \bibfield  {author} {\bibinfo {author} {\bibfnamefont {M.}~\bibnamefont
  {Otteneder}}, \bibinfo {author} {\bibfnamefont {S.}~\bibnamefont {Hubmann}},
  \bibinfo {author} {\bibfnamefont {X.}~\bibnamefont {Lu}}, \bibinfo {author}
  {\bibfnamefont {D.~A.}\ \bibnamefont {Kozlov}}, \bibinfo {author}
  {\bibfnamefont {L.~E.}\ \bibnamefont {Golub}}, \bibinfo {author}
  {\bibfnamefont {K.}~\bibnamefont {Watanabe}}, \bibinfo {author}
  {\bibfnamefont {T.}~\bibnamefont {Taniguchi}}, \bibinfo {author}
  {\bibfnamefont {D.~K.}\ \bibnamefont {Efetov}},\ and\ \bibinfo {author}
  {\bibfnamefont {S.~D.}\ \bibnamefont {Ganichev}},\ }\bibfield  {title}
  {\bibinfo {title} {Terahertz photogalvanics in twisted bilayer graphene close
  to the second magic angle},\ }\href@noop {} {\bibfield  {journal} {\bibinfo
  {journal} {Nano Letters}\ }\textbf {\bibinfo {volume} {20}},\ \bibinfo
  {pages} {7152} (\bibinfo {year} {2020})}\BibitemShut {NoStop}%
\bibitem [{\citenamefont {Hubmann}\ \emph {et~al.}(2022)\citenamefont
  {Hubmann}, \citenamefont {Soul}, \citenamefont {Di~Battista}, \citenamefont
  {Hild}, \citenamefont {Watanabe}, \citenamefont {Taniguchi}, \citenamefont
  {Efetov},\ and\ \citenamefont {Ganichev}}]{Hubmann22}%
  \BibitemOpen
  \bibfield  {author} {\bibinfo {author} {\bibfnamefont {S.}~\bibnamefont
  {Hubmann}}, \bibinfo {author} {\bibfnamefont {P.}~\bibnamefont {Soul}},
  \bibinfo {author} {\bibfnamefont {G.}~\bibnamefont {Di~Battista}}, \bibinfo
  {author} {\bibfnamefont {M.}~\bibnamefont {Hild}}, \bibinfo {author}
  {\bibfnamefont {K.}~\bibnamefont {Watanabe}}, \bibinfo {author}
  {\bibfnamefont {T.}~\bibnamefont {Taniguchi}}, \bibinfo {author}
  {\bibfnamefont {D.~K.}\ \bibnamefont {Efetov}},\ and\ \bibinfo {author}
  {\bibfnamefont {S.~D.}\ \bibnamefont {Ganichev}},\ }\bibfield  {title}
  {\bibinfo {title} {Nonlinear intensity dependence of photogalvanics and
  photoconductance induced by terahertz laser radiation in twisted bilayer
  graphene close to magic angle},\ }\href
  {https://doi.org/10.1103/PhysRevMaterials.6.024003} {\bibfield  {journal}
  {\bibinfo  {journal} {Phys. Rev. Mater.}\ }\textbf {\bibinfo {volume} {6}},\
  \bibinfo {pages} {024003} (\bibinfo {year} {2022})}\BibitemShut {NoStop}%
\bibitem [{\citenamefont {Ma}\ \emph {et~al.}(2022)\citenamefont {Ma},
  \citenamefont {Yuan}, \citenamefont {Cheung}, \citenamefont {Watanabe},
  \citenamefont {Taniguchi}, \citenamefont {Zhang},\ and\ \citenamefont
  {Xia}}]{Ma22}%
  \BibitemOpen
  \bibfield  {author} {\bibinfo {author} {\bibfnamefont {C.}~\bibnamefont
  {Ma}}, \bibinfo {author} {\bibfnamefont {S.}~\bibnamefont {Yuan}}, \bibinfo
  {author} {\bibfnamefont {P.}~\bibnamefont {Cheung}}, \bibinfo {author}
  {\bibfnamefont {K.}~\bibnamefont {Watanabe}}, \bibinfo {author}
  {\bibfnamefont {T.}~\bibnamefont {Taniguchi}}, \bibinfo {author}
  {\bibfnamefont {F.}~\bibnamefont {Zhang}},\ and\ \bibinfo {author}
  {\bibfnamefont {F.}~\bibnamefont {Xia}},\ }\bibfield  {title} {\bibinfo
  {title} {Intelligent infrared sensing enabled by tunable moir{\'e} quantum
  geometry},\ }\href@noop {} {\bibfield  {journal} {\bibinfo  {journal}
  {Nature}\ }\textbf {\bibinfo {volume} {604}},\ \bibinfo {pages} {266}
  (\bibinfo {year} {2022})}\BibitemShut {NoStop}%
\bibitem [{\citenamefont {Lopes~dos Santos}\ \emph {et~al.}(2012)\citenamefont
  {Lopes~dos Santos}, \citenamefont {Peres},\ and\ \citenamefont
  {Castro~Neto}}]{dosSantos12}%
  \BibitemOpen
  \bibfield  {author} {\bibinfo {author} {\bibfnamefont {J.~M.~B.}\
  \bibnamefont {Lopes~dos Santos}}, \bibinfo {author} {\bibfnamefont
  {N.~M.~R.}\ \bibnamefont {Peres}},\ and\ \bibinfo {author} {\bibfnamefont
  {A.~H.}\ \bibnamefont {Castro~Neto}},\ }\bibfield  {title} {\bibinfo {title}
  {Continuum model of the twisted graphene bilayer},\ }\href
  {https://doi.org/10.1103/PhysRevB.86.155449} {\bibfield  {journal} {\bibinfo
  {journal} {Phys. Rev. B}\ }\textbf {\bibinfo {volume} {86}},\ \bibinfo
  {pages} {155449} (\bibinfo {year} {2012})}\BibitemShut {NoStop}%
\bibitem [{\citenamefont {Aversa}\ and\ \citenamefont {Sipe}(1995)}]{Aversa95}%
  \BibitemOpen
  \bibfield  {author} {\bibinfo {author} {\bibfnamefont {C.}~\bibnamefont
  {Aversa}}\ and\ \bibinfo {author} {\bibfnamefont {J.~E.}\ \bibnamefont
  {Sipe}},\ }\bibfield  {title} {\bibinfo {title} {Nonlinear optical
  susceptibilities of semiconductors: Results with a length-gauge analysis},\
  }\href {https://doi.org/10.1103/PhysRevB.52.14636} {\bibfield  {journal}
  {\bibinfo  {journal} {Phys. Rev. B}\ }\textbf {\bibinfo {volume} {52}},\
  \bibinfo {pages} {14636} (\bibinfo {year} {1995})}\BibitemShut {NoStop}%
\bibitem [{\citenamefont {Sipe}\ and\ \citenamefont
  {Shkrebtii}(2000)}]{Shkrebtii00}%
  \BibitemOpen
  \bibfield  {author} {\bibinfo {author} {\bibfnamefont {J.~E.}\ \bibnamefont
  {Sipe}}\ and\ \bibinfo {author} {\bibfnamefont {A.~I.}\ \bibnamefont
  {Shkrebtii}},\ }\bibfield  {title} {\bibinfo {title} {Second-order optical
  response in semiconductors},\ }\href
  {https://doi.org/10.1103/PhysRevB.61.5337} {\bibfield  {journal} {\bibinfo
  {journal} {Phys. Rev. B}\ }\textbf {\bibinfo {volume} {61}},\ \bibinfo
  {pages} {5337} (\bibinfo {year} {2000})}\BibitemShut {NoStop}%
\bibitem [{SM()}]{SM}%
  \BibitemOpen
  \href@noop {} {}\bibinfo {note} {See Supplemental Material at XXX for a
  summary of the length gauge formalism in real and momentum space, an extended
  symmetry analysis, and an extended discussion on the constraints due to
  particle-hole symmetry\red{, and a calculation of the injection current LPGE in the absence of time-reversal symmetry}.}\BibitemShut {Stop}%
\bibitem [{\citenamefont {Moon}\ and\ \citenamefont {Koshino}(2013)}]{Moon13}%
  \BibitemOpen
  \bibfield  {author} {\bibinfo {author} {\bibfnamefont {P.}~\bibnamefont
  {Moon}}\ and\ \bibinfo {author} {\bibfnamefont {M.}~\bibnamefont {Koshino}},\
  }\bibfield  {title} {\bibinfo {title} {Optical absorption in twisted bilayer
  graphene},\ }\href {https://doi.org/10.1103/PhysRevB.87.205404} {\bibfield
  {journal} {\bibinfo  {journal} {Phys. Rev. B}\ }\textbf {\bibinfo {volume}
  {87}},\ \bibinfo {pages} {205404} (\bibinfo {year} {2013})}\BibitemShut
  {NoStop}%
\bibitem [{\citenamefont {Ahn}\ and\ \citenamefont {Nagaosa}(2021)}]{Ahn21}%
  \BibitemOpen
  \bibfield  {author} {\bibinfo {author} {\bibfnamefont {J.}~\bibnamefont
  {Ahn}}\ and\ \bibinfo {author} {\bibfnamefont {N.}~\bibnamefont {Nagaosa}},\
  }\bibfield  {title} {\bibinfo {title} {Theory of optical responses in clean
  multi-band superconductors},\ }\href@noop {} {\bibfield  {journal} {\bibinfo
  {journal} {Nature communications}\ }\textbf {\bibinfo {volume} {12}},\
  \bibinfo {pages} {1617} (\bibinfo {year} {2021})}\BibitemShut {NoStop}%
\bibitem [{\citenamefont {Song}\ \emph {et~al.}(2019)\citenamefont {Song},
  \citenamefont {Wang}, \citenamefont {Shi}, \citenamefont {Li}, \citenamefont
  {Fang},\ and\ \citenamefont {Bernevig}}]{Song19}%
  \BibitemOpen
  \bibfield  {author} {\bibinfo {author} {\bibfnamefont {Z.}~\bibnamefont
  {Song}}, \bibinfo {author} {\bibfnamefont {Z.}~\bibnamefont {Wang}}, \bibinfo
  {author} {\bibfnamefont {W.}~\bibnamefont {Shi}}, \bibinfo {author}
  {\bibfnamefont {G.}~\bibnamefont {Li}}, \bibinfo {author} {\bibfnamefont
  {C.}~\bibnamefont {Fang}},\ and\ \bibinfo {author} {\bibfnamefont {B.~A.}\
  \bibnamefont {Bernevig}},\ }\bibfield  {title} {\bibinfo {title} {All magic
  angles in twisted bilayer graphene are topological},\ }\href
  {https://doi.org/10.1103/PhysRevLett.123.036401} {\bibfield  {journal}
  {\bibinfo  {journal} {Phys. Rev. Lett.}\ }\textbf {\bibinfo {volume} {123}},\
  \bibinfo {pages} {036401} (\bibinfo {year} {2019})}\BibitemShut {NoStop}%
\bibitem [{\citenamefont {Song}\ \emph {et~al.}(2021)\citenamefont {Song},
  \citenamefont {Lian}, \citenamefont {Regnault},\ and\ \citenamefont
  {Bernevig}}]{Song21}%
  \BibitemOpen
  \bibfield  {author} {\bibinfo {author} {\bibfnamefont {Z.-D.}\ \bibnamefont
  {Song}}, \bibinfo {author} {\bibfnamefont {B.}~\bibnamefont {Lian}}, \bibinfo
  {author} {\bibfnamefont {N.}~\bibnamefont {Regnault}},\ and\ \bibinfo
  {author} {\bibfnamefont {B.~A.}\ \bibnamefont {Bernevig}},\ }\bibfield
  {title} {\bibinfo {title} {Twisted bilayer graphene. ii. stable symmetry
  anomaly},\ }\href {https://doi.org/10.1103/PhysRevB.103.205412} {\bibfield
  {journal} {\bibinfo  {journal} {Phys. Rev. B}\ }\textbf {\bibinfo {volume}
  {103}},\ \bibinfo {pages} {205412} (\bibinfo {year} {2021})}\BibitemShut
  {NoStop}%
\bibitem [{\citenamefont {Bultinck}\ \emph {et~al.}(2020)\citenamefont
  {Bultinck}, \citenamefont {Khalaf}, \citenamefont {Liu}, \citenamefont
  {Chatterjee}, \citenamefont {Vishwanath},\ and\ \citenamefont
  {Zaletel}}]{Bultinck20}%
  \BibitemOpen
  \bibfield  {author} {\bibinfo {author} {\bibfnamefont {N.}~\bibnamefont
  {Bultinck}}, \bibinfo {author} {\bibfnamefont {E.}~\bibnamefont {Khalaf}},
  \bibinfo {author} {\bibfnamefont {S.}~\bibnamefont {Liu}}, \bibinfo {author}
  {\bibfnamefont {S.}~\bibnamefont {Chatterjee}}, \bibinfo {author}
  {\bibfnamefont {A.}~\bibnamefont {Vishwanath}},\ and\ \bibinfo {author}
  {\bibfnamefont {M.~P.}\ \bibnamefont {Zaletel}},\ }\bibfield  {title}
  {\bibinfo {title} {Ground state and hidden symmetry of magic-angle graphene
  at even integer filling},\ }\href
  {https://doi.org/10.1103/PhysRevX.10.031034} {\bibfield  {journal} {\bibinfo
  {journal} {Phys. Rev. X}\ }\textbf {\bibinfo {volume} {10}},\ \bibinfo
  {pages} {031034} (\bibinfo {year} {2020})}\BibitemShut {NoStop}%
\bibitem [{\citenamefont {Bernevig}\ \emph {et~al.}(2021)\citenamefont
  {Bernevig}, \citenamefont {Song}, \citenamefont {Regnault},\ and\
  \citenamefont {Lian}}]{Bernevig21}%
  \BibitemOpen
  \bibfield  {author} {\bibinfo {author} {\bibfnamefont {B.~A.}\ \bibnamefont
  {Bernevig}}, \bibinfo {author} {\bibfnamefont {Z.-D.}\ \bibnamefont {Song}},
  \bibinfo {author} {\bibfnamefont {N.}~\bibnamefont {Regnault}},\ and\
  \bibinfo {author} {\bibfnamefont {B.}~\bibnamefont {Lian}},\ }\bibfield
  {title} {\bibinfo {title} {Twisted bilayer graphene. iii. interacting
  hamiltonian and exact symmetries},\ }\href
  {https://doi.org/10.1103/PhysRevB.103.205413} {\bibfield  {journal} {\bibinfo
   {journal} {Phys. Rev. B}\ }\textbf {\bibinfo {volume} {103}},\ \bibinfo
  {pages} {205413} (\bibinfo {year} {2021})}\BibitemShut {NoStop}%
\bibitem [{\citenamefont {Sauer}\ \emph {et~al.}(2023)\citenamefont {Sauer},
  \citenamefont {Taghizadeh}, \citenamefont {Petralanda}, \citenamefont
  {Ovesen}, \citenamefont {Thygesen}, \citenamefont {Olsen}, \citenamefont
  {Cornean},\ and\ \citenamefont {Pedersen}}]{Sauer23}%
  \BibitemOpen
  \bibfield  {author} {\bibinfo {author} {\bibfnamefont {M.~O.}\ \bibnamefont
  {Sauer}}, \bibinfo {author} {\bibfnamefont {A.}~\bibnamefont {Taghizadeh}},
  \bibinfo {author} {\bibfnamefont {U.}~\bibnamefont {Petralanda}}, \bibinfo
  {author} {\bibfnamefont {M.}~\bibnamefont {Ovesen}}, \bibinfo {author}
  {\bibfnamefont {K.~S.}\ \bibnamefont {Thygesen}}, \bibinfo {author}
  {\bibfnamefont {T.}~\bibnamefont {Olsen}}, \bibinfo {author} {\bibfnamefont
  {H.}~\bibnamefont {Cornean}},\ and\ \bibinfo {author} {\bibfnamefont {T.~G.}\
  \bibnamefont {Pedersen}},\ }\bibfield  {title} {\bibinfo {title} {Shift
  current photovoltaic efficiency of 2d materials},\ }\href@noop {} {\bibfield
  {journal} {\bibinfo  {journal} {npj Computational Materials}\ }\textbf
  {\bibinfo {volume} {9}},\ \bibinfo {pages} {35} (\bibinfo {year}
  {2023})}\BibitemShut {NoStop}%
\bibitem [{\citenamefont {Monteverde}\ \emph {et~al.}(2010)\citenamefont
  {Monteverde}, \citenamefont {Ojeda-Aristizabal}, \citenamefont {Weil},
  \citenamefont {Bennaceur}, \citenamefont {Ferrier}, \citenamefont {Gu\'eron},
  \citenamefont {Glattli}, \citenamefont {Bouchiat}, \citenamefont {Fuchs},\
  and\ \citenamefont {Maslov}}]{Monteverde10}%
  \BibitemOpen
  \bibfield  {author} {\bibinfo {author} {\bibfnamefont {M.}~\bibnamefont
  {Monteverde}}, \bibinfo {author} {\bibfnamefont {C.}~\bibnamefont
  {Ojeda-Aristizabal}}, \bibinfo {author} {\bibfnamefont {R.}~\bibnamefont
  {Weil}}, \bibinfo {author} {\bibfnamefont {K.}~\bibnamefont {Bennaceur}},
  \bibinfo {author} {\bibfnamefont {M.}~\bibnamefont {Ferrier}}, \bibinfo
  {author} {\bibfnamefont {S.}~\bibnamefont {Gu\'eron}}, \bibinfo {author}
  {\bibfnamefont {C.}~\bibnamefont {Glattli}}, \bibinfo {author} {\bibfnamefont
  {H.}~\bibnamefont {Bouchiat}}, \bibinfo {author} {\bibfnamefont {J.~N.}\
  \bibnamefont {Fuchs}},\ and\ \bibinfo {author} {\bibfnamefont {D.~L.}\
  \bibnamefont {Maslov}},\ }\bibfield  {title} {\bibinfo {title} {Transport and
  elastic scattering times as probes of the nature of impurity scattering in
  single-layer and bilayer graphene},\ }\href
  {https://doi.org/10.1103/PhysRevLett.104.126801} {\bibfield  {journal}
  {\bibinfo  {journal} {Phys. Rev. Lett.}\ }\textbf {\bibinfo {volume} {104}},\
  \bibinfo {pages} {126801} (\bibinfo {year} {2010})}\BibitemShut {NoStop}%
\bibitem [{\citenamefont {Hejazi}\ \emph {et~al.}(2019)\citenamefont {Hejazi},
  \citenamefont {Liu}, \citenamefont {Shapourian}, \citenamefont {Chen},\ and\
  \citenamefont {Balents}}]{Hejazi19}%
  \BibitemOpen
  \bibfield  {author} {\bibinfo {author} {\bibfnamefont {K.}~\bibnamefont
  {Hejazi}}, \bibinfo {author} {\bibfnamefont {C.}~\bibnamefont {Liu}},
  \bibinfo {author} {\bibfnamefont {H.}~\bibnamefont {Shapourian}}, \bibinfo
  {author} {\bibfnamefont {X.}~\bibnamefont {Chen}},\ and\ \bibinfo {author}
  {\bibfnamefont {L.}~\bibnamefont {Balents}},\ }\bibfield  {title} {\bibinfo
  {title} {Multiple topological transitions in twisted bilayer graphene near
  the first magic angle},\ }\href {https://doi.org/10.1103/PhysRevB.99.035111}
  {\bibfield  {journal} {\bibinfo  {journal} {Phys. Rev. B}\ }\textbf {\bibinfo
  {volume} {99}},\ \bibinfo {pages} {035111} (\bibinfo {year}
  {2019})}\BibitemShut {NoStop}%
\bibitem [{\citenamefont {Tan}\ and\ \citenamefont {Rappe}(2016)}]{Tan16}%
  \BibitemOpen
  \bibfield  {author} {\bibinfo {author} {\bibfnamefont {L.~Z.}\ \bibnamefont
  {Tan}}\ and\ \bibinfo {author} {\bibfnamefont {A.~M.}\ \bibnamefont
  {Rappe}},\ }\bibfield  {title} {\bibinfo {title} {Enhancement of the bulk
  photovoltaic effect in topological insulators},\ }\href
  {https://doi.org/10.1103/PhysRevLett.116.237402} {\bibfield  {journal}
  {\bibinfo  {journal} {Phys. Rev. Lett.}\ }\textbf {\bibinfo {volume} {116}},\
  \bibinfo {pages} {237402} (\bibinfo {year} {2016})}\BibitemShut {NoStop}%
\bibitem [{\citenamefont {Cook}\ \emph {et~al.}(2017)\citenamefont {Cook},
  \citenamefont {M.~Fregoso}, \citenamefont {De~Juan}, \citenamefont {Coh},\
  and\ \citenamefont {Moore}}]{Cook17}%
  \BibitemOpen
  \bibfield  {author} {\bibinfo {author} {\bibfnamefont {A.~M.}\ \bibnamefont
  {Cook}}, \bibinfo {author} {\bibfnamefont {B.}~\bibnamefont {M.~Fregoso}},
  \bibinfo {author} {\bibfnamefont {F.}~\bibnamefont {De~Juan}}, \bibinfo
  {author} {\bibfnamefont {S.}~\bibnamefont {Coh}},\ and\ \bibinfo {author}
  {\bibfnamefont {J.~E.}\ \bibnamefont {Moore}},\ }\bibfield  {title} {\bibinfo
  {title} {Design principles for shift current photovoltaics},\ }\href@noop {}
  {\bibfield  {journal} {\bibinfo  {journal} {Nat. Commun.}\ }\textbf {\bibinfo
  {volume} {8}},\ \bibinfo {pages} {14176} (\bibinfo {year}
  {2017})}\BibitemShut {NoStop}%
\bibitem [{\citenamefont {Yan}(2018)}]{Yan18}%
  \BibitemOpen
  \bibfield  {author} {\bibinfo {author} {\bibfnamefont {Z.}~\bibnamefont
  {Yan}},\ }\bibfield  {title} {\bibinfo {title} {Precise determination of
  critical points of topological phase transitions via shift current in
  two-dimensional inversion asymmetric insulators},\ }\href@noop {} {\bibfield
  {journal} {\bibinfo  {journal} {arXiv preprint arXiv:1812.02191}\ } (\bibinfo
  {year} {2018})}\BibitemShut {NoStop}%
\bibitem [{\citenamefont {Sivianes}\ and\ \citenamefont {Iba\~nez
  Azpiroz}(2023)}]{Sivianes23}%
  \BibitemOpen
  \bibfield  {author} {\bibinfo {author} {\bibfnamefont {J.}~\bibnamefont
  {Sivianes}}\ and\ \bibinfo {author} {\bibfnamefont {J.}~\bibnamefont
  {Iba\~nez-Azpiroz}},\ }\bibfield  {title} {\bibinfo {title} {Shift
  photoconductivity in the haldane model},\ }\href
  {https://doi.org/10.1103/PhysRevB.108.155419} {\bibfield  {journal} {\bibinfo
   {journal} {Phys. Rev. B}\ }\textbf {\bibinfo {volume} {108}},\ \bibinfo
  {pages} {155419} (\bibinfo {year} {2023})}\BibitemShut {NoStop}%
\bibitem [{\citenamefont {Vafek}\ and\ \citenamefont {Kang}(2020)}]{Vafek20RG}%
  \BibitemOpen
  \bibfield  {author} {\bibinfo {author} {\bibfnamefont {O.}~\bibnamefont
  {Vafek}}\ and\ \bibinfo {author} {\bibfnamefont {J.}~\bibnamefont {Kang}},\
  }\bibfield  {title} {\bibinfo {title} {Renormalization group study of hidden
  symmetry in twisted bilayer graphene with coulomb interactions},\ }\href
  {https://doi.org/10.1103/PhysRevLett.125.257602} {\bibfield  {journal}
  {\bibinfo  {journal} {Phys. Rev. Lett.}\ }\textbf {\bibinfo {volume} {125}},\
  \bibinfo {pages} {257602} (\bibinfo {year} {2020})}\BibitemShut {NoStop}%
\bibitem [{\citenamefont {Kang}\ and\ \citenamefont
  {Vafek}(2023)}]{KangVafekPH23}%
  \BibitemOpen
  \bibfield  {author} {\bibinfo {author} {\bibfnamefont {J.}~\bibnamefont
  {Kang}}\ and\ \bibinfo {author} {\bibfnamefont {O.}~\bibnamefont {Vafek}},\
  }\bibfield  {title} {\bibinfo {title} {Pseudomagnetic fields, particle-hole
  asymmetry, and microscopic effective continuum hamiltonians of twisted
  bilayer graphene},\ }\href {https://doi.org/10.1103/PhysRevB.107.075408}
  {\bibfield  {journal} {\bibinfo  {journal} {Phys. Rev. B}\ }\textbf {\bibinfo
  {volume} {107}},\ \bibinfo {pages} {075408} (\bibinfo {year}
  {2023})}\BibitemShut {NoStop}%
\bibitem [{\citenamefont {Scheer}\ \emph {et~al.}(2022)\citenamefont {Scheer},
  \citenamefont {Gu},\ and\ \citenamefont {Lian}}]{Scheer22}%
  \BibitemOpen
  \bibfield  {author} {\bibinfo {author} {\bibfnamefont {M.~G.}\ \bibnamefont
  {Scheer}}, \bibinfo {author} {\bibfnamefont {K.}~\bibnamefont {Gu}},\ and\
  \bibinfo {author} {\bibfnamefont {B.}~\bibnamefont {Lian}},\ }\bibfield
  {title} {\bibinfo {title} {Magic angles in twisted bilayer graphene near
  commensuration: Towards a hypermagic regime},\ }\href
  {https://doi.org/10.1103/PhysRevB.106.115418} {\bibfield  {journal} {\bibinfo
   {journal} {Phys. Rev. B}\ }\textbf {\bibinfo {volume} {106}},\ \bibinfo
  {pages} {115418} (\bibinfo {year} {2022})}\BibitemShut {NoStop}%
\bibitem [{\citenamefont {Guinea}\ and\ \citenamefont
  {Walet}(2019)}]{Guinea19}%
  \BibitemOpen
  \bibfield  {author} {\bibinfo {author} {\bibfnamefont {F.}~\bibnamefont
  {Guinea}}\ and\ \bibinfo {author} {\bibfnamefont {N.~R.}\ \bibnamefont
  {Walet}},\ }\bibfield  {title} {\bibinfo {title} {Continuum models for
  twisted bilayer graphene: Effect of lattice deformation and hopping
  parameters},\ }\href {https://doi.org/10.1103/PhysRevB.99.205134} {\bibfield
  {journal} {\bibinfo  {journal} {Phys. Rev. B}\ }\textbf {\bibinfo {volume}
  {99}},\ \bibinfo {pages} {205134} (\bibinfo {year} {2019})}\BibitemShut
  {NoStop}%
\bibitem [{\citenamefont {Fang}\ \emph {et~al.}(2019)\citenamefont {Fang},
  \citenamefont {Carr}, \citenamefont {Zhu}, \citenamefont {Massatt},\ and\
  \citenamefont {Kaxiras}}]{Fang19}%
  \BibitemOpen
  \bibfield  {author} {\bibinfo {author} {\bibfnamefont {S.}~\bibnamefont
  {Fang}}, \bibinfo {author} {\bibfnamefont {S.}~\bibnamefont {Carr}}, \bibinfo
  {author} {\bibfnamefont {Z.}~\bibnamefont {Zhu}}, \bibinfo {author}
  {\bibfnamefont {D.}~\bibnamefont {Massatt}},\ and\ \bibinfo {author}
  {\bibfnamefont {E.}~\bibnamefont {Kaxiras}},\ }\bibfield  {title} {\bibinfo
  {title} {Angle-dependent $\{$$\backslash$it Ab initio$\}$ low-energy
  hamiltonians for a relaxed twisted bilayer graphene heterostructure},\
  }\href@noop {} {\bibfield  {journal} {\bibinfo  {journal} {arXiv:1908.00058}\
  } (\bibinfo {year} {2019})}\BibitemShut {NoStop}%
\bibitem [{\citenamefont {Koshino}\ and\ \citenamefont
  {Nam}(2020)}]{Koshino20}%
  \BibitemOpen
  \bibfield  {author} {\bibinfo {author} {\bibfnamefont {M.}~\bibnamefont
  {Koshino}}\ and\ \bibinfo {author} {\bibfnamefont {N.~N.~T.}\ \bibnamefont
  {Nam}},\ }\bibfield  {title} {\bibinfo {title} {Effective continuum model for
  relaxed twisted bilayer graphene and moir\'e electron-phonon interaction},\
  }\href {https://doi.org/10.1103/PhysRevB.101.195425} {\bibfield  {journal}
  {\bibinfo  {journal} {Phys. Rev. B}\ }\textbf {\bibinfo {volume} {101}},\
  \bibinfo {pages} {195425} (\bibinfo {year} {2020})}\BibitemShut {NoStop}%
\bibitem [{\citenamefont {Gao}\ \emph {et~al.}(2020)\citenamefont {Gao},
  \citenamefont {Zhang},\ and\ \citenamefont {Xiao}}]{Gao20}%
  \BibitemOpen
  \bibfield  {author} {\bibinfo {author} {\bibfnamefont {Y.}~\bibnamefont
  {Gao}}, \bibinfo {author} {\bibfnamefont {Y.}~\bibnamefont {Zhang}},\ and\
  \bibinfo {author} {\bibfnamefont {D.}~\bibnamefont {Xiao}},\ }\bibfield
  {title} {\bibinfo {title} {Tunable layer circular photogalvanic effect in
  twisted bilayers},\ }\href {https://doi.org/10.1103/PhysRevLett.124.077401}
  {\bibfield  {journal} {\bibinfo  {journal} {Phys. Rev. Lett.}\ }\textbf
  {\bibinfo {volume} {124}},\ \bibinfo {pages} {077401} (\bibinfo {year}
  {2020})}\BibitemShut {NoStop}%
\bibitem [{\citenamefont {Matsyshyn}\ \emph {et~al.}(2023)\citenamefont
  {Matsyshyn}, \citenamefont {Xiong}, \citenamefont {Arora},\ and\
  \citenamefont {Song}}]{Matsyshyn23}%
  \BibitemOpen
  \bibfield  {author} {\bibinfo {author} {\bibfnamefont {O.}~\bibnamefont
  {Matsyshyn}}, \bibinfo {author} {\bibfnamefont {Y.}~\bibnamefont {Xiong}},
  \bibinfo {author} {\bibfnamefont {A.}~\bibnamefont {Arora}},\ and\ \bibinfo
  {author} {\bibfnamefont {J.~C.~W.}\ \bibnamefont {Song}},\ }\bibfield
  {title} {\bibinfo {title} {Layer photovoltaic effect in van der waals
  heterostructures},\ }\href {https://doi.org/10.1103/PhysRevB.107.205306}
  {\bibfield  {journal} {\bibinfo  {journal} {Phys. Rev. B}\ }\textbf {\bibinfo
  {volume} {107}},\ \bibinfo {pages} {205306} (\bibinfo {year}
  {2023})}\BibitemShut {NoStop}%
\bibitem [{\citenamefont {Blount}(1962)}]{Blount}%
  \BibitemOpen
  \bibfield  {author} {\bibinfo {author} {\bibfnamefont {E.}~\bibnamefont
  {Blount}},\ }\bibfield  {title} {\bibinfo {title} {Formalisms of band
  theory},\ }\bibfield  {booktitle} {\emph {\bibinfo {booktitle} {Solid state
  physics}},\ }\href@noop {} {\ \textbf {\bibinfo {volume} {13}},\ \bibinfo
  {pages} {305} (\bibinfo {year} {1962})}\BibitemShut {NoStop}%
\bibitem [{\citenamefont {Ventura}\ \emph {et~al.}(2017)\citenamefont
  {Ventura}, \citenamefont {Passos}, \citenamefont {Lopes~dos Santos},
  \citenamefont {Viana Parente~Lopes},\ and\ \citenamefont
  {Peres}}]{Ventura17}%
  \BibitemOpen
  \bibfield  {author} {\bibinfo {author} {\bibfnamefont {G.~B.}\ \bibnamefont
  {Ventura}}, \bibinfo {author} {\bibfnamefont {D.~J.}\ \bibnamefont {Passos}},
  \bibinfo {author} {\bibfnamefont {J.~M.~B.}\ \bibnamefont {Lopes~dos
  Santos}}, \bibinfo {author} {\bibfnamefont {J.~M.}\ \bibnamefont {Viana
  Parente~Lopes}},\ and\ \bibinfo {author} {\bibfnamefont {N.~M.~R.}\
  \bibnamefont {Peres}},\ }\bibfield  {title} {\bibinfo {title} {Gauge
  covariances and nonlinear optical responses},\ }\href
  {https://doi.org/10.1103/PhysRevB.96.035431} {\bibfield  {journal} {\bibinfo
  {journal} {Phys. Rev. B}\ }\textbf {\bibinfo {volume} {96}},\ \bibinfo
  {pages} {035431} (\bibinfo {year} {2017})}\BibitemShut {NoStop}%
\bibitem [{\citenamefont {de~Juan}\ \emph {et~al.}(2020)\citenamefont
  {de~Juan}, \citenamefont {Zhang}, \citenamefont {Morimoto}, \citenamefont
  {Sun}, \citenamefont {Moore},\ and\ \citenamefont {Grushin}}]{deJuan20}%
  \BibitemOpen
  \bibfield  {author} {\bibinfo {author} {\bibfnamefont {F.}~\bibnamefont
  {de~Juan}}, \bibinfo {author} {\bibfnamefont {Y.}~\bibnamefont {Zhang}},
  \bibinfo {author} {\bibfnamefont {T.}~\bibnamefont {Morimoto}}, \bibinfo
  {author} {\bibfnamefont {Y.}~\bibnamefont {Sun}}, \bibinfo {author}
  {\bibfnamefont {J.~E.}\ \bibnamefont {Moore}},\ and\ \bibinfo {author}
  {\bibfnamefont {A.~G.}\ \bibnamefont {Grushin}},\ }\bibfield  {title}
  {\bibinfo {title} {Difference frequency generation in topological
  semimetals},\ }\href {https://doi.org/10.1103/PhysRevResearch.2.012017}
  {\bibfield  {journal} {\bibinfo  {journal} {Phys. Rev. Res.}\ }\textbf
  {\bibinfo {volume} {2}},\ \bibinfo {pages} {012017(R)} (\bibinfo {year}
  {2020})}\BibitemShut {NoStop}%
\bibitem [{\citenamefont {Golub}\ \emph {et~al.}(2011)\citenamefont {Golub},
  \citenamefont {Tarasenko}, \citenamefont {Entin},\ and\ \citenamefont
  {Magarill}}]{Golub11}%
  \BibitemOpen
  \bibfield  {author} {\bibinfo {author} {\bibfnamefont {L.~E.}\ \bibnamefont
  {Golub}}, \bibinfo {author} {\bibfnamefont {S.~A.}\ \bibnamefont
  {Tarasenko}}, \bibinfo {author} {\bibfnamefont {M.~V.}\ \bibnamefont
  {Entin}},\ and\ \bibinfo {author} {\bibfnamefont {L.~I.}\ \bibnamefont
  {Magarill}},\ }\bibfield  {title} {\bibinfo {title} {Valley separation in
  graphene by polarized light},\ }\href
  {https://doi.org/10.1103/PhysRevB.84.195408} {\bibfield  {journal} {\bibinfo
  {journal} {Phys. Rev. B}\ }\textbf {\bibinfo {volume} {84}},\ \bibinfo
  {pages} {195408} (\bibinfo {year} {2011})}\BibitemShut {NoStop}%
\bibitem [{\citenamefont {Ahn}\ and\ \citenamefont {Ghosh}(2023)}]{Ahn23}%
  \BibitemOpen
  \bibfield  {author} {\bibinfo {author} {\bibfnamefont {J.}~\bibnamefont
  {Ahn}}\ and\ \bibinfo {author} {\bibfnamefont {B.}~\bibnamefont {Ghosh}},\
  }\bibfield  {title} {\bibinfo {title} {Topological circular dichroism in
  chiral multifold semimetals},\ }\href
  {https://doi.org/10.1103/PhysRevLett.131.116603} {\bibfield  {journal}
  {\bibinfo  {journal} {Phys. Rev. Lett.}\ }\textbf {\bibinfo {volume} {131}},\
  \bibinfo {pages} {116603} (\bibinfo {year} {2023})}\BibitemShut {NoStop}%
\end{thebibliography}
\end{document}


\title{Supplemental Material: Intrinsic and extrinsic photogalvanic effects in twisted bilayer graphene}
\author{Fernando Peñaranda}
\affiliation{Donostia International Physics Center, P. Manuel de Lardizabal 4, 20018 Donostia-San Sebastian, Spain}
\author{Héctor Ochoa}
\affiliation{Department of Physics, Columbia University, New York, NY 10027, USA}
\author{Fernando de Juan}
\affiliation{Donostia International Physics Center, P. Manuel de Lardizabal 4, 20018 Donostia-San Sebastian, Spain}
\affiliation{IKERBASQUE, Basque Foundation for Science, Maria Diaz de Haro 3, 48013 Bilbao, Spain}
\date{\today}
\maketitle
\appendix
\onecolumngrid
\vspace{-2em}
\tableofcontents

\section{Optical response in the length gauge}

In this section we summarize the derivation of the photogalvanic tensors in the length gauge~\cite{Aversa95,Shkrebtii00} for a quasi-two dimensional material which is periodic in the in-plane directions but finite in the $z$ direction. The coupling to light in the length gauge ($\hbar=1$) is 
\begin{align}
H = H_0 - e\vec E \cdot \vec r    
\end{align}
where $\vec E$ is the time-dependent electric field, $\vec r$ is the position operator, and $H_0$ defines the eigenstate basis $H_0\left|n\right> = \omega_n \left|n\right>$.  The current operator $\vec J = e \vec v$ is written in terms of the velocity $\vec v = \partial_t \vec r = -i [H,\vec r]$ and its expectation value is computed in terms of the density matrix $\rho$ as
\begin{align}
\vec J = e \sum_{nm} \vec v_{nm} \rho_{mn} \label{current}
\end{align}
where $\vec v_{nm}=\left<n|\vec v|m\right>$ and the density matrix satisfies the dynamical equation $i \partial_t \rho = [H,\rho]$, which in matrix elements reads 
\begin{align}
i \partial_t \rho_{nm} = \omega_{nm} \rho_{nm} - \vec E\cdot [\vec r,\rho]_{nm}  \label{dynamical}
\end{align}
$\rho_{nm}$ can be obtained in powers of $\vec E$ by iterating this equation, starting from the equilibrium value $\rho_{nm}^{(0)} = f_n-f_m$ with $f_n$ the Fermi occupation, and substitution in Eq. \ref{current} gives the current to any desired order.  

In the presence of translational invariance, these equations can be Fourier transformed to momentum space following the standard approach~\cite{Aversa95,Shkrebtii00}. For a quasi 2D material, only in the in-plane coordinates are transformed, so for the rest of this section we will use latin indices $i=x,y$ to denote spatial components, and write the $z$ index explicitly as $\vec r \cdot \vec E = r^i E^i + r^z E^z$. We only consider currents $J^i$ that flow within the plane. Out-of-plane currents, or interlayer currents, cannot be measured in a standard photocurrent experiment. Light-induced static polarization, as opposed to current, can be produced in the out of plane direction \cite{Gao20,Matsyshyn23} but is not the focus of this work. 

The apparent complication of calculations involving the position operator matrix elements in momentum space is that they are not diagonal in $k$ and singular~\cite{Blount}. However, if the position operator appears in a commutator with another operator $S$ that is diagonal in $k$ then the commutator matrix elements are diagonal and well behaved,
\begin{align}
\left<n,\vec k|[r^a,S] |m, \vec k'\right>&= i D_a S_{nm}\delta(\vec k- \vec k'),
\end{align}
where $D_a$ is covariant derivative acting on the matrix $S_{nm}$,
\begin{align}
D_a S_{nm} = \partial_{k_a} S_{nm} -i \sum_p(\mathcal{A}_{np}S_{pm}-S_{np}\mathcal{A}_{pm})\label{fullCovariantD},
\end{align}
and $\mathcal{A}_{nm}^a = i\left<n|\partial_{k_a}|m\right>$ is the Berry connection matrix~\cite{Ventura17}. Because of this property, the current only contains matrix elements that are diagonal in $k$,  
\begin{align}
J^i = e \int \frac{d^2k}{(2\pi)^2} \sum_{nm} v^i_{mn} \rho_{nm},
\end{align}
where $v^i_{nm} = D_a S_{nm}$, and momentum labels for the eigenstates are dropped for clarity. Similarly, the dynamical equation is 
\begin{align}
(i \partial_t -\omega_{nm})\rho_{nm} = - i\,E^i D_i \rho_{nm} -E^z [r^z,\rho]_{nm} .
\label{dynrhomomentum}
\end{align}

Next, we follow Refs. ~\cite{Aversa95,Shkrebtii00} and divide the current operator into diagonal and off-diagonal components,
\begin{align}
J^i = J^i_o+J^i_d = \partial_t \vec r_o + [H,r^i_d], \label{SipeCurrent}
\end{align}
where the time derivative is kept explicit in the first term, but left as a commutator in the second. The commutator of $r_d^i$ can be written in terms of the diagonal covariant derivative, denoted as 
\begin{align}
[r^i_d,S]_{nm} = iS_{nm;i} = i[\partial_{k_i}S_{nm} - i(\mathcal{A}_{nn}-\mathcal{A}_{mm})S_{nm}],   
\end{align}
leading to 
\begin{align}
J_o^i &= \left< \partial_t r_o^i \right> = \partial_t \sum_{n \neq m}  \mathcal{A}^i_{nm} \rho_{nm} ,\label{Jinter} \\
J_d^i &=\left< [H, r_d^i] \right> = \left< [H_0,r_d^i] - E^j[ r_o^j+r_d^j, r_d^i]-E^z[r^z, r_d^i]\right>  ,
\end{align}
and the diagonal current evaluates to
\begin{align}
J_d^i= \sum_{nm} \left(\delta_{nm} \partial_i \omega_n - E^j \left[A_{nm;i}^j - \delta_{nm} (\partial_{k_i}A^j_n - \partial_{k_j}A^i_n)-E^z r^z_{nm;i}\right]\right) \rho_{nm} .\label{Jintra}
\end{align}

Finally, for a monochromatic wave $\vec E(t) = \vec E_\beta e^{i \omega_\beta t}$ we collect all the terms to second order in $\vec E$ and zero total frequency (see the Supplemental Material in Ref. \cite{deJuan20} for a detailed derivation) and find
\begin{align}
 J^{\text{inj}}_i &= 2 \eta_{ijk} E_j E_k^* \label{injectionSM},\\
J^{\text{shift}}_i &= 2 \sigma_{ijk} E_j E_k^*,
\end{align}
with
\begin{align}
\sigma_{ijk} = \frac{\pi e^3}{2\hbar^2} \int \frac{d^2k}{(2\pi)^2} \sum_{nm} f_{nm}[r^k_{nm;i}r^j_{mn}-r^k_{nm}r^j_{mn;i}] \delta(\omega - \omega_{nm}), \\
\sigma_{ijz} = \frac{\pi e^3}{2\hbar^2} \int \frac{d^2k}{(2\pi)^2} \sum_{nm}  f_{nm}[r^z_{nm;i}r^j_{mn}-r^z_{nm}r^j_{mn;i}] \delta(\omega - \omega_{nm}) ,
\end{align}
for the shift currents and 
\begin{align}
\eta_{ijk} = \tau \frac{\pi e^3}{\hbar^2} \int \frac{d^2k}{(2\pi)^2} \sum_{nm}  f_{nm}\partial_{k_i}\omega_{nm}r^j_{nm}r^k_{mn} \delta(\omega - \omega_{nm}),\\ 
\eta_{izk} = \tau \frac{\pi e^3}{\hbar^2} \int \frac{d^2k}{(2\pi)^2} \sum_{nm}  f_{nm}\partial_{k_i}\omega_{nm}r^z_{nm}r^k_{mn} \delta(\omega - \omega_{nm}) 
\end{align}
for the injection currents. In the relaxation time approximation, for the injection current we consider its saturated value so Eq. \ref{injectionSM} turns into $J^{\text{inj}}_i = \eta_{ijk}E_jE_k^*$ as used in the main text. In the presence of time-reversal symmetry, $\sigma_{ijk}$ becomes symmetric in the last two indices, while $\eta_{ijk}$ becomes antisymmetric, producing Eq. 1 in the main text. 

The generalized derivatives require the evaluation of wavefunction derivatives which cannot be performed numerically in a straightforward way, so we resort to the usual sum rule for in-plane components

\begin{align}
r^a_{nm;b}(k) = -\frac{1}{i\omega_{nm}} \left[ \frac{v_{nm}^a \Delta_{mn}^b + v_{nm}^b\Delta_{mn}^a}{\omega_{nm}} - w^{ba}_{nm} + \sum_{p \neq,n,m} \frac{v^a_{np}v^b_{pm}}{\omega_{pm}}-\frac{v^b_{np}v^a_{pm}}{\omega_{np}} \right].
\end{align}
where $w^{ba}_{nm} = \langle n| \partial_{k_b} \partial_{k_a} H|m\rangle$~\cite{Cook17}, which is zero in our model.
For the out of plane position operator, we can obtain a new sum rule from $[r^z,r^i] =0$, which gives 
\begin{align}
[r^z,r_d^i+r_o^i]_{nm} = ir^z_{nm;i} + \sum_p r^z_{np}A^i_{pm} - A^i_{np}r^z_{pm} =0   ,
\end{align}
and using $i \omega_{nm} A^i_{nm} = v^i_{nm}$ for $n \neq m$ we arrive at
\begin{align}
r^z_{nm;i} &= \sum_{p\neq m} r^z_{np}\frac{v^i_{pm}}{\omega_{pm}} -  \sum_{p\neq n} \frac{v^i_{np}}{\omega_{np}}r^z_{pm} .
\end{align}

\section{Extended symmetry analysis}

In the general case without time-reversal symmetry, both shift and injection currents contribute to both LPGE, $\sigma_{ijk}$, and CPGE, $\eta_{ijk}$. Their general expressions can be written as
\begin{align}
J_i^{\rm shift} &= \sigma_{ijk}^{\rm shift}(E_jE_k^* + E_j^*E_k) + \eta_{ijk}^{\rm shift}(E_jE_k^* - E_j^*E_k) \\
J_i^{\rm inj} &= \sigma_{ijk}^{\rm inj}(E_jE_k^* + E_j^*E_k) + \eta_{ijk}^{\rm inj}(E_jE_k^* -E_j^*E_k) 
\end{align}
where $\sigma_{abc}^{\rm shift}$ is real, symmetric (LPGE) and $\mathcal{T}$-even, $\eta_{abc}^{\rm shift}$ is imaginary, antisymmetric (CPGE) and $\mathcal{T}$-odd, $\eta_{abc}^{\rm inj}$ is imaginary, antisymmetric (CPGE) and $\mathcal{T}$-even, and $\sigma_{abc}^{\rm inj}$ is real, symmetric (LPGE) and $\mathcal{T}$-odd. Their expressions are explicitly
\begin{align}
\sigma_{ijk}^{\rm shift} &= \frac{\pi e^3}{2\hbar^2} \int_{\boldsymbol{k}} \sum_{nm} f_{nm}{\rm Re}[r^k_{nm;i}r^j_{mn}-r^k_{nm}r^j_{mn;i}] \delta(\omega - \omega_{nm}),\\
\eta_{ijk}^{\rm shift} &= \frac{\pi e^3}{2\hbar^2} \int_{\boldsymbol{k}} \sum_{nm} f_{nm}{\rm Im}[r^k_{nm;i}r^j_{mn}-r^k_{nm}r^j_{mn;i}] \delta(\omega - \omega_{nm}),\\
\eta_{ijk}^{\rm inj} &= \tau \frac{\pi e^3}{\hbar^2}\int_{\boldsymbol{k}} \sum_{nm} f_{nm} \partial_{k_i}\omega_{nm} {\rm Im}[r_{nm}^jr_{mn}^k] \delta(\omega - \omega_{nm}), \\
\sigma_{ijk}^{\rm inj} &= \tau \frac{\pi e^3}{\hbar^2}\int_{\boldsymbol{k}} \sum_{nm} f_{nm} \partial_{k_i}\omega_{nm} {\rm Re}[r_{nm}^jr_{mn}^k] \delta(\omega - \omega_{nm}).  
\end{align}

In the presence of time-reversal symmetry, the $\mathcal{T}$-odd parts vanish, $\sigma_{abc}^{\rm inj}=0$ and $\eta_{abc}^{\rm shift}=0$ so in the main text we define for simplicity 
\begin{align}
\sigma_{ijk} & \equiv \sigma_{ijk}^{\rm shift} \;\; ({\rm main \; text})\\
\eta_{ijk} & \equiv \eta_{ijk}^{\rm inj} \;\; \;\;({\rm main \; text})
\end{align}
and consider only those components. In the presence of the combined $\mathcal{T}*I$ symmetry, with $I$ the inversion operator, the $\mathcal{T}$-even parts vanish, $\eta_{abc}^{\rm inj}=0$ and $\sigma_{abc}^{\rm shift}=0$. If there is any other point group element which would make a certain tensor component to vanish, for example $C_{2z}$ for the in-plane components, then the presence of $C_{2z}\mathcal{T}$ makes the $\mathcal{T}$-even in-plane components vanish as well. The way these time-reversal constraints for shift and injection currents are implemented in the TBG continuum model is non trivial because of the presence of valley symmetry and the fact that $\mathcal{T}$ interchanges valleys. Within a single valley the model has magnetic point group $6'22'$ generated by $C_{3z}$, $C_{2x}$ and $C_{2z}*\mathcal{T}$. Even in the absence of symmetry breaking, the reduced point group symmetry in a single valley would allow $\sigma_{xxx} = -\sigma_{xyy} = -\sigma_{yxy}$ in addition to the intrinsic ones, and $C_{2z}*\mathcal{T}$ further requires that LPGE originates only from the $\mathcal{T}$-odd injection current $\sigma_{ijk}^{\rm inj}$ \cite{Kaplan22}. However this component must vanish in the physical two valley model due to $C_{2z}$, so that this is actually a pure valley photocurrent \cite{Golub11}. In the presence of lattice symmetry breaking, further contributions to such valley-odd photocurrents can also be obtained.  For the intrinsic component $\sigma_{xyz}$, $C_{2z}*\mathcal{T}$ allows only the $\mathcal{T}$-even shift current $\sigma_{xyz}^{\rm shift}$ in the one valley model. For CPGE, $\eta_{xyz}$ only has contributions from $\mathcal{T}$-even injection $\eta_{xyz}^{\rm inj}$ in the one valley model. 

\begin{figure}[H]
    \centering
\includegraphics[width=1\textwidth]{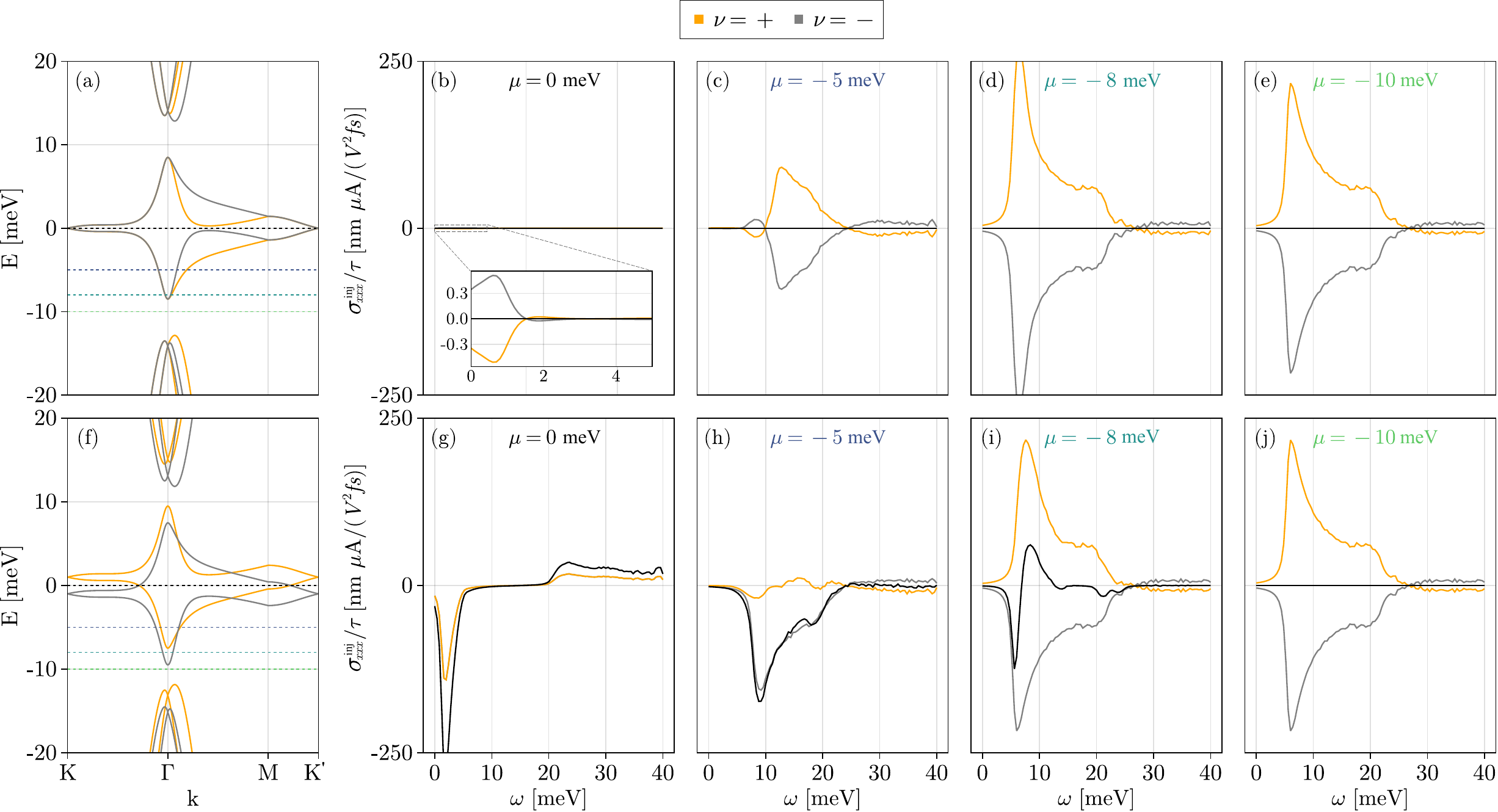}
    \caption{In-plane $xxx$ injection current component in the presence (first row) and absence (second row) of time-reversal symmetry. Time reversal symmetry is broken by a valley-polarized perturbation $\Delta_4 \rho_z$. (a) Bandstructures for $\Delta_4 =0$ and (b-e)
   corresponding frequency dependency of $\sigma^{\rm inj}_{xxx}/\tau$ at different $\mu$ values corresponding to the dashed lines in (a). (f) Bandstructures for $\Delta_4 =1$ meV and (g-j) corresponding $\sigma^{\rm inj}_{xxx}/\tau$. The contributions of the positive and negative valleys are depicted in orange and grey, respectively, and their sum in black. $\theta = 1^\circ$.}
    \label{S4}
\end{figure}

For completeness, here we report the calculation of the $\mathcal{T}$-odd valley photocurrent $\sigma_{xxx}^{\rm inj}$, which can be found in Fig. \ref{S4}. The upper row shows the valley degenerate case, where the photocurrents for each valley are exactly opopsite and cancel in the total response, for any frequency and chemical potential. The lower row shows a case with finite valley polarization, where the total response is generally finite. 

\section{Particle-hole symmetry}

Twisted bilayer graphene has an approximate particle-hole symmetry $\mathcal{C}$: $U_C H U_C = -H^*$ with $U_C = \sigma_x \tau_y$; here $\tau_y$ is a Pauli matrix acting on layers.
In the continuum model, this symmetry is exact if we neglect the relative rotation of the Pauli matrices on each layer and the Kang-Vafek term $w_3$.

Particle-hole symmetry requires that if $\left|n,k\right>$ is an eigenstate with energy $E_n$, then the state  $\left|-n,-k\right> = C\left|n,k\right>$ is an eigenstate with momentum $-k$ and energy $-E_n$ (labeled by $-n$). This is shown by acting with $H_{-k}$ on this state 
\begin{align}
H_{-k} C \left|n,k\right> &= H_{-k} U_C \left|n,k\right>^* = U_C (-H_{k}^*) \left|n,k\right>^* = - U_C (H_{k} \left|n,k\right> )^* \nonumber\\
&= -E_n C \left|n,k\right>
\end{align}

This symmetry implies the following constraints on matrix elements. Consider an Hermitian operator $O$ with matrix element between states $n$ and $m$ given by $O_{nm}(k) = \left<n,k|O(k)|m,k\right>$. Inserting the squared particle-hole operator we find
\begin{align}
O_{nm}(k) &= - \left<n,k|O(k) C^2 |m,k\right> \\
&= - \left<n,k|O(k) U_C (|-m,-k\right>)^*  \\
&= - \left<-m,-k| U_C^T O^T(k)  (|n,k\right>)^* \\
&= \left<-m,-k| U_C O^*(k) U_C  |-n,-k\right>,
\end{align}
where we have used that $\left<n|A|m\right> = \psi_{n,i}^* A_{ij} \psi_j = \psi_j A^T_{ji} \psi_{n,i}^* = \left<n|^*A^T|m\right>^*$. For the in-plane position matrix element $O = i\partial_{k_i}$  we have explicitly
\begin{align}
r^i_{nm}(k) = \left<-m,-k| -i\partial_{k_i}  |-n,-k\right> = r^i_{-m-n}(-k)  ,
\end{align}
while for the out of plane position $r^z=\tau_z$
\begin{align}
r^z_{nm}(k) = -\left<-m,-k| \tau_z  |-n,-k\right> = - r^z_{-m-n}(-k) .
\end{align}
The diagonal velocity satisfies 
\begin{align}
v^i_{nn}(k) = \partial_{k_i}E_n(k)  = \partial_{-k_i}[E_{-n}(-k)] = v^i_{-n-n}(-k),
\end{align}
while the off-diagonal one satisfies
\begin{align}
v^i_{nm}(k) = i\omega_{nm}(k) r_{nm}^i(k) = i\omega_{-m-n}(-k) r_{-m-n}^i(-k) = v^i_{-m-n}(-k).
\end{align}
Finally, the generalized derivative is 
\begin{align}
r^i_{nm;j}(k) &= [\partial_{k_j} -iA_{nn}(k) + iA_{mm}(k)]r^i_{nm}(k) \nonumber \\
&= [-\partial_{-k_j} -iA_{-n-n}(-k) + iA_{-m-m}(-k)]r^i_{-m-n}(-k) = -r^i_{-m-n;j}(-k).
\end{align}

In the more restrictive case where $k$ maps to itself under $C$ and we consider the matrix element of $O$ between conjugate states, we have
\begin{align}
\left<n,k|O(k)|-n,k\right> &= \left<n,k|O(k)U_C|n,k\right>^* \nonumber\\
&= \left<n,k|U_C U_C O(k) U_C|n,k\right>^* \nonumber\\
&= -\left<n,k|U_C O^T(k)U_C|-n,k\right>. \label{PHconjugates}
\end{align}
For example, for the velocity operator $v^i = \partial_{k_i}H_k$ we have $U_C \partial_{k_i}H_k U_C = - \partial_{k_i}H_{-k} =\partial_{-k_i}H_{-k}$. When $k=-k$ up to a reciprocal lattice vector, this implies from Eq. \ref{PHconjugates} that $v_{n-n}(k) = - v_{n-n}(k)$, which reproduces the result in Ref. \cite{Moon13} that the transitions at the M point between conjugate bands have zero velocity matrix element.

With these transformation properties we can now rearrange the integrals for optical responses as follows. All integrals are of the type
\begin{align}
I = \int d^2k \sum_{n>0,m<0} F(k,n,m)  \delta(\omega_{nm}-\omega) ,   
\end{align}
where $n$ runs over empty states while $m$ runs over occupied states. For every pair $n,m$ at momentum $k$ there is another pair $-m,-n$ at momentum $-k$ with an optical transition related by particle hole symmetry since $\omega_{nm} = \omega_{-m-n}$. Therefore, we can rewrite the integral as
\begin{align}
I = \int_{k>0} d^2k \sum_{n>0,m<0} [F(k,n,m) + F(-k,-m,-n)]  \delta(\omega_{nm}-\omega).    
\end{align}
The constraints are 
\begin{itemize}
    \item For the optical conductivity, $F= r_{nm}^i(k) r_{mn}^j(k)$ and $F(-k,-m,-n)= r_{-m-n}^i(-k) r_{-n-m}^j(-k) = r_{nm}^i(k) r_{mn}^j(k) = F(k,n,m)$, so there is no constraint from particle-hole symmetry. 
    \item For the shift current with only in-plane components, $F= r_{nm;j}^i(k) r_{mn}^k(k)$ and $F(-k,-m,-n)= -F(k,n,m)$, so the shift current should vanish for $\mu=0$.
    \item For the shift current with one out of plane component, $F= r_{nm;j}^i(k) r_{mn}^k(k)$ and $F(-k,-m,-n)= F(k,n,m)$, so there is no constraint. 
    \item For the injection current with in-plane components, $F= (v^k_{nn}-v^k_{mm})r_{nm}^i(k) r_{mn}^k(k)$ and $F(-k,-m,-n)= (v^k_{-m-m}(-k)-v^k_{-n-n}(-k))r_{-m-n}^i(-k) r_{-n-m}^k(-k) = - F(k,n,m)$, so it should also vanish for $\mu=0$. Similarly as with the shift current, if there is a single out-of-plane component then there is no constraint. 
    \item Finally, we can consider circular dichroism, where Ref. \cite{Ahn23} gives the explicit interband expression with $F = r^a_{nm}r^c_{mp}v^b_{pn} - r^b_{nm}r^c_{mp}v^a_{pn} -(v^c_{nn}+v^c_{mm})r_{nm}^a r_{mn}^b$. Comparison with the injection current reveals that when all components are in-plane, there is no constraint. But when there is one out of plane component (the usual case where $J_x = \sigma_{xyz}E_y q_z$) then the circular dichroism vanishes due to particle-hole, as stated in \cite{Morell17}. 
    
\end{itemize}

All of these constraints are shown in the main text as Eqs.  7-10, in the presence of chemical potential $\mu$, which is odd under PHS.  

The breaking of PHS in the continuum model with local tunneling comes both from the rotation of the Pauli matrices and the $w_3$ term. However, we can use a unitary rotation $U= e^{i \alpha \sigma_z \tau_z}$ which leaves the $w_{AB}$ tunneling term invariant and eliminates the rotation of the Pauli matrices while modifying the numerical value of $w_3$~\cite{Scheer22}. Since both in-plane and out of plane position operators are invariant under this transformation, for all computations of optical responses we can disregard the rotation of Pauli matrices and only consider an effective value of $w_3$. 

\begin{figure}[H]
    \centering
\includegraphics[width=1\textwidth]{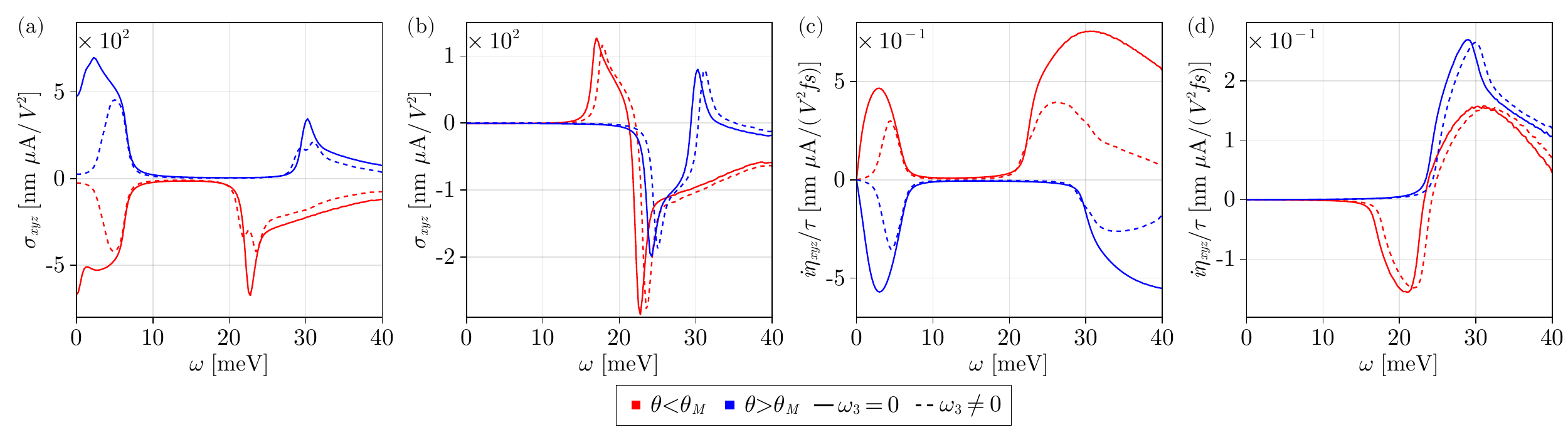}
    \caption{Effect of the $\omega_3$ term on the intrinsic photogalvanic tensors at off-normal incidence. (a) Shift current $\sigma_{xyz}$ at neutrality ($\mu = \mu_1$ in Fig. 1c). (b) $\sigma_{xyz}$ for fully filled flat bands ($\mu = \mu_2$). (c,d) Same for the injection current $\eta_{xyz}$. Red (blue) encodes $\theta<\theta_M$ ($\theta>\theta_M$), while solid and dashed lines refer to $\omega_3 = 0$ and $\omega_3 = 0.9$ meV, respectively.}
    \label{S1}
\end{figure}

\begin{figure}[H]
    \centering
\includegraphics[width=1\textwidth]{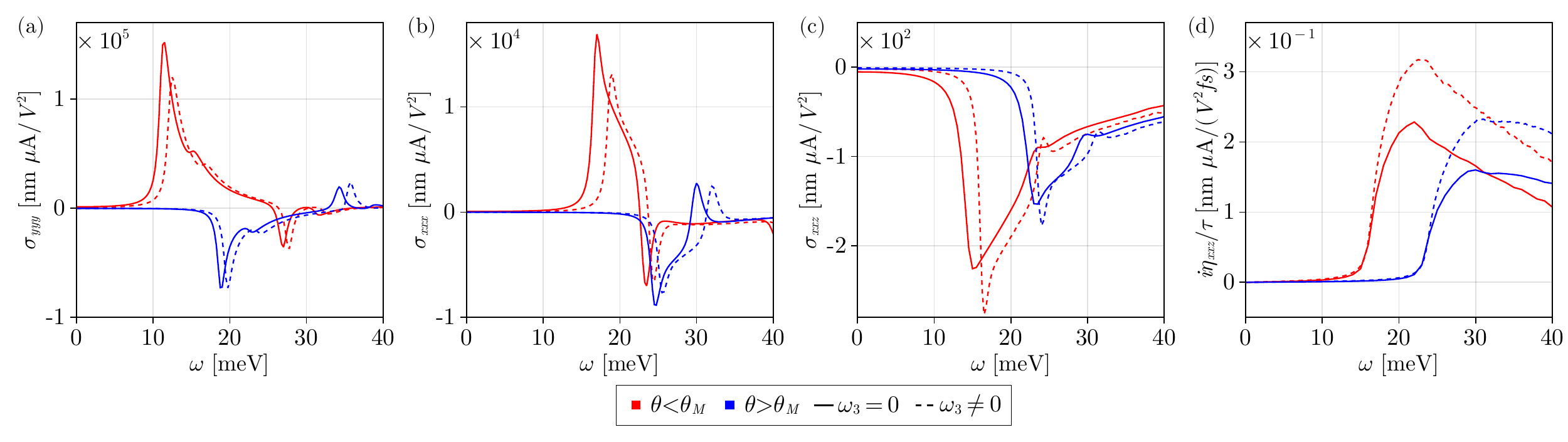}
    \caption{Effect of the $\omega_3$ term on the photogalvanic components induced by symmetry-breaking perturbations: (a) $\sigma_{yyy}$ with a $\Delta_1 \sigma_z$ term, (b) $\sigma_{xxx}$ with a $\Delta_2 \sigma_z \tau_z$ term, and (c) $\sigma_{xxz}$ and (d) $\eta_{xxz}$ with a $\Delta_3 \tau_z$ term. Red (blue) encodes $\theta<\theta_M$ ($\theta>\theta_M$), while solid and dashed lines refer to $\omega_3 = 0$ and $\omega_3 = 0.9$ meV, respectively. $\mu = -\mu_2$ in all panels (see Fig.1c). The rest of parameters are the same as for Fig. 3.}
    \label{S2}
\end{figure}

To substantiate the claim that the breaking of PHS only leads to minor modifications of the photocurrents which are already finite in the presence of PHS, here we reproduce the calculations in the main text in the presence of $w_3=0.9$ meV. Figures \ref{S1} and \ref{S2} show the intrinsic and extrinsic photogalvanic tensors in Figs. 2 and 3 of the main text, respectively, at $\theta = 1.05^\circ<\theta_M$ (red) and $\theta = 1.12^\circ>\theta_M$ (blue). Across all panels, a finite $\omega_3$ value (dashed) leads to a small quantitative modulation with respect to the $\omega_3 = 0$ responses shown as solid lines.

\section{Twist angle dependency of the intrinsic photocurrents}

\begin{figure}[h!]
    \centering
\includegraphics[width=.7\textwidth]{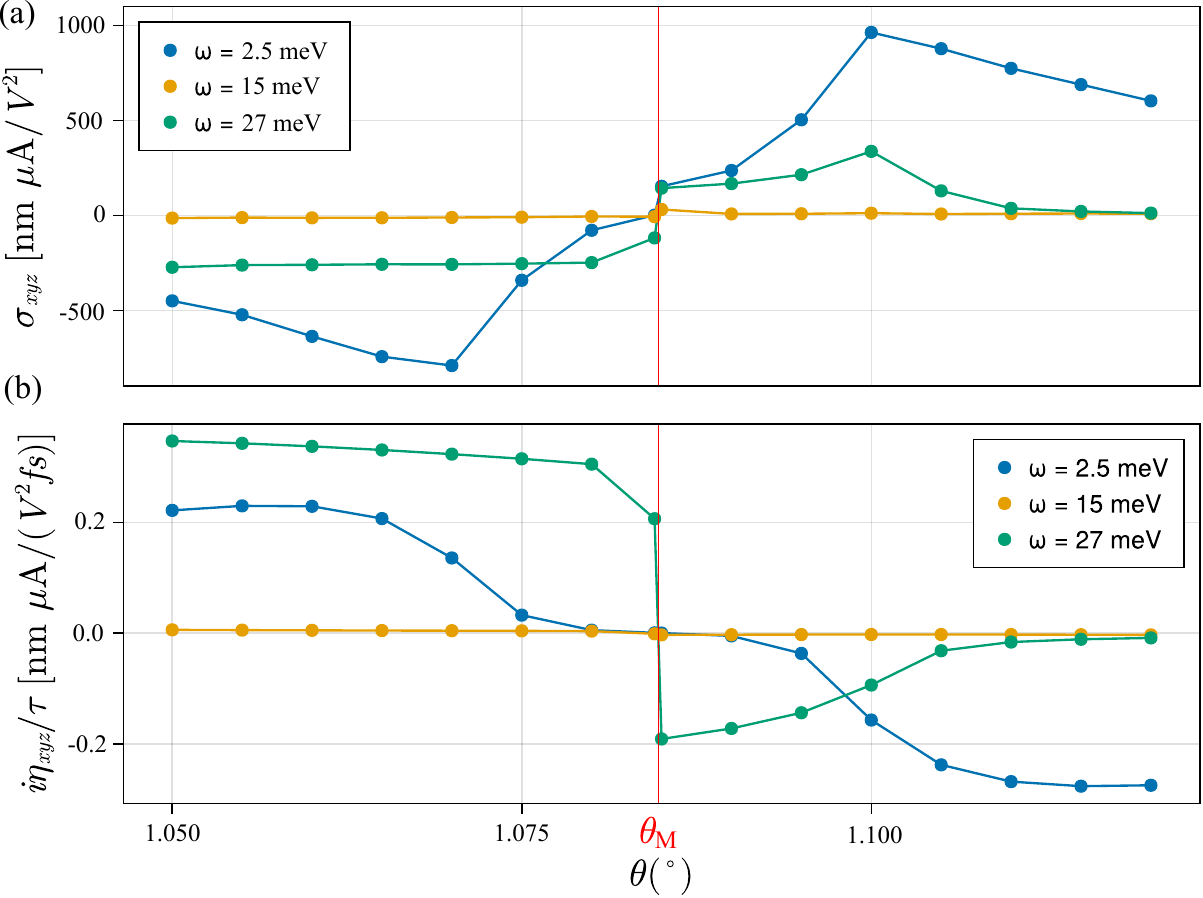}
    \caption{Intrinsic photogalvanic tensors at off-normal incidence as a function of the twist angle. At neutrality (a) $\sigma_{xyz}(\theta)$ and (b) $i \eta_{xyz}/\tau(\theta)$  are shown for three different values of $\omega = \{2.5, 15, 27\}$ meV in blue, yellow, and green, respectively. The red vertical line denotes the magic angle at which these components change sign. Same parameters as in Fig. 2 of the main text.}
    \label{S3}
\end{figure}

In the main text we have shown how the intrinsic photocurrents flip their sign at the magic angle. Here, in Fig. \ref{S3}, we present complementary plots of this inversion for fixed frequency as a function of twist angle, where the effect is also appreciated. 

%